\documentclass[acmsmall]{acmart}

\usepackage{lineno}   
\usepackage{booktabs} 
\usepackage{hyperref}   
\usepackage{url}  

\usepackage{multirow}  
\usepackage{color}  
\newtheorem{strategy}{Strategy}    
\newtheorem{upper bound}{Upper bound}

\usepackage{diagbox}
\usepackage{amsmath}
\usepackage{arydshln} 
\usepackage{color} 

\usepackage[linesnumbered,ruled]{algorithm2e} 

\SetAlFnt{\small}
\SetAlCapFnt{\small}
\SetAlCapNameFnt{\small}
\SetAlCapHSkip{0pt}
\IncMargin{-\parindent}

\acmDOI{0000001.0000001}
\begin{document}


\title{TALENT: Targeted Mining of Non-overlapping Sequential Patterns}

\author{Zefeng Chen}
\affiliation{ 
	\institution{Jinan University}
	\city{Guangzhou}
	\country{China}
}
\email{czf1027@gmail.com}

\author{Wensheng Gan}
\authornote{This is the corresponding author}
\affiliation{
	\institution{Jinan University}
	\city{Guangzhou}
	\country{China}
}
\email{wsgan001@gmail.com}

\author{Gengsen Huang}
\affiliation{
	\institution{Jinan University}
	\city{Guangzhou}
	\country{China}
}
\email{hgengsen@gmail.com}

\author{Zhenlian Qi}
\affiliation{%
    \institution{Guangdong Eco-Engineering Polytechnic}
    \city{Guangzhou}
    \country{China}
}
\email{qzlhit@gmail.com}

\author{Yan Li}
\affiliation{%
    \institution{Hebei University of Technology}
    \city{Tianjin}
    \country{China}
}
\email{lywuc@163.com}

\author{Philip S. Yu}
\affiliation{
	\institution{University of Illinois Chicago}
	\city{Chicago}
	\country{USA}
}
\email{psyu@uic.edu}

\begin{abstract}
    With the widespread application of efficient pattern mining algorithms, sequential patterns that allow gap constraints have become a valuable tool to discover knowledge from biological data such as DNA and protein sequences. Among all kinds of gap-constrained mining, non-overlapping sequence mining can mine interesting patterns and satisfy the anti-monotonic property (the Apriori property). However, existing algorithms do not search for targeted sequential patterns, resulting in unnecessary and redundant pattern generation. Targeted pattern mining can not only mine patterns that are more interesting to users but also reduce the unnecessary redundant sequence generated, which can greatly avoid irrelevant computation. In this paper, we define and formalize the problem of targeted non-overlapping sequential pattern mining and propose an algorithm named TALENT (\underline{TA}rgeted mining of Sequentia\underline{L} Patt\underline{E}r\underline{N} with Cons\underline{T}raints). Two search methods including breadth-first and depth-first searching are designed to troubleshoot the generation of patterns. Furthermore, several pruning strategies to reduce the reading of sequences and items in the data and terminate redundant pattern extensions are presented. Finally, we select a series of datasets with different characteristics and conduct extensive experiments to compare the TALENT algorithm with the existing algorithms for mining non-overlapping sequential patterns. The experimental results demonstrate that the proposed targeted mining algorithm, TALENT, has excellent mining efficiency and can deal efficiently with many different query settings.
\end{abstract}

%
%



%
%


\maketitle

\renewcommand{\shortauthors}{Z. Chen \textit{et al.}}

\section{Introduction}

With the booming development of the Internet, a massive amount of data inundated our lives and become an indispensable part \cite{gan2017data}. Under this circumstance, mining information that users are interested in from massive data has become a very significant topic. Among the technologies for mining information, pattern mining is an essential field of study for discovering interesting patterns in databases \cite{fournier2014spmf}. As a branch of pattern mining, frequent pattern mining (FPM) \cite{han2007frequent} is a mature topic that aims to mine patterns that appear frequently in databases, as patterns that occur more often are more probable to have features that interest users. The mining goal of FPM is to discover patterns whose occurrences are greater than the minimum support threshold \textit{minsup} set by users. However, FPM ignores the order in which items occur, resulting in some valuable patterns not being found or meaningless patterns being found. In order to address this problem, sequential pattern mining (SPM) \cite{fournier2017survey,gan2019survey} was presented and applied by many researchers. Owing to the consideration of sequential order, SPM can mine more meaningful and valuable patterns than FPM. For example, in FPM, patterns $<$A, B$>$ and $<$B, A$>$ are the same, and both can be uniformly donated as $<$A, B$>$. However, in SPM, they are absolutely different. In many domains, the order of items is of great importance. When analyzing text, the order of words in a sentence and sentences in a paragraph is required to be considered. In DNA sequence analysis, the order of deoxy nucleotides cannot be ignored \cite{wu2017nosep}. Furthermore, the order of events also plays a key role in cyberattack detection \cite{tianfield2017data}.

In some application fields such as biology, medicine, and pharmacy, the algorithms of pattern mining are required to mine varied biological data \cite{li2012efficient} such as DNA sequences, RNA sequences, amino acid sequences in peptide chains. Because of the possibility of mutations in DNA or RNA genes, users may be required to mine frequent sequential patterns while allowing additional items to occur between items in the pattern when processing these data. These requirements are also in the transcription and translation errors in the proteins. These additional items can be represented by gap constraints. For example, the pattern $<$A, C, T$>$ in the original DNA sequence "AGCAT" is allowed to have gap constraints. Formally, according to the research \cite{wu2021ntp}, the wildcard `?' is used to indicate the match of a single character. While `*' is used to indicate the match of multiple characters. For example, $<$A, ?, C$>$ can match with $<$A, G, C$>$ and $<$A, T, C$>$. And for $<$A, *, C$>$, it can match $<$A, G, C$>$, $<$A, G, G, C$>$, $<$A, T, C$>$, and $<$A, T, T, C$>$, etc. In recent years, the gap constraints have been defined to illustrate the wildcard of characters between characters in patterns \cite{wu2017nosep}. A pattern with gap constraints is described as $P$ = $<p_1$, $[a_1,b_1]$, $p_2$, $[a_2,b_2]$, $p_3$, ..., $p_{k-1}$, $[a_{k-1}$, $b_{k-1}]$, $p_k>$, where $a_j$ and $b_j$ represent the minimum number and the maximum number of the wildcard `?' between characters $p_i$ and $p_{i+1}$, respectively. For example, $<$T, [0, 2], C$>$ can represent three cases of gap constraints, which are $<$T, C$>$, $<$T, ?, C$>$, and $<$T, ?, ?, C$>$ expressed in the form of wildcards `?'. When $a_1$ = $a_2$ = ... = $a_{k-1}$ and $b_1$ = $b_2$ = ... = $b_{k-1}$, it can be abbreviated as $P$ = $<p_1, p_2, p_3, $..., $p_{k-1}, p_k>$ with gap constraints \textit{gap} = [\textit{mingap}, \textit{maxgap}], where \textit{mingap} = $a_1$ = $a_2$ = ... = $a_{k-1}$ and \textit{maxgap} = $b_1$ = $b_2$ = ... = $b_{k-1}$. This formulation is clear and has been widely used by many researchers and many application fields, such as behavioral sequence analysis \cite{de2019mining}, keyword extraction analysis \cite{xie2017efficient}, and biological sequence analysis \cite{wu2017nosep}.

Nonetheless, the current research on gap-constrained FPM does not focus on the query sequence of interest to users. For example, if a medical professional needs to investigate the content of a DNA fragment in a disease, only the extension sequences of a specific DNA fragment are most likely to be associated. However, the mined results contain a lot of uninteresting sequences, which causes a lot of waste. In order to filter out information and results that are not useful to the user, targeted pattern mining provides an efficient solution to the classic pattern mining problem \cite{fournier2022pattern}. For example, given a sequence $S$ = $s_1s_2s_3s_4s_5s_6s_7s_8$ = "ACTACTGA" without considering the gap and length constraints, and the query sequence set by the user is $<$T, A$>$ and \textit{minsup} = 1. Although the occurrences of $<$A, C, T$>$ are larger than \textit{minsup}, it is not considered a frequent target pattern. However, because $<$T, A$>$ is a sub-pattern of $<$T, G, A$>$ and the occurrence of $<$T, G, A$>$ is 1 $\ge$ \textit{minsup}, it is a target frequent pattern in this mining task. Obviously, using traditional SPM algorithms to mine these frequent target patterns is inefficient. Many patterns that do not include the query sequence are also generated \cite{huang2022taspm}. This would cause a lot of irrelevant computations. Therefore, targeted pattern mining is proposed to solve this problem. Specifically, utilizing the idea of target queries, the user could input one query sequence (or more in some papers) at a time. The efficient algorithms can discover the desired patterns that include the input query sequence, and thus there is no need to discover and extend numerous patterns that are hopeless to extend to frequent target patterns. 

\begin{figure}[h]
    \centering
    \includegraphics[clip,scale=0.2]{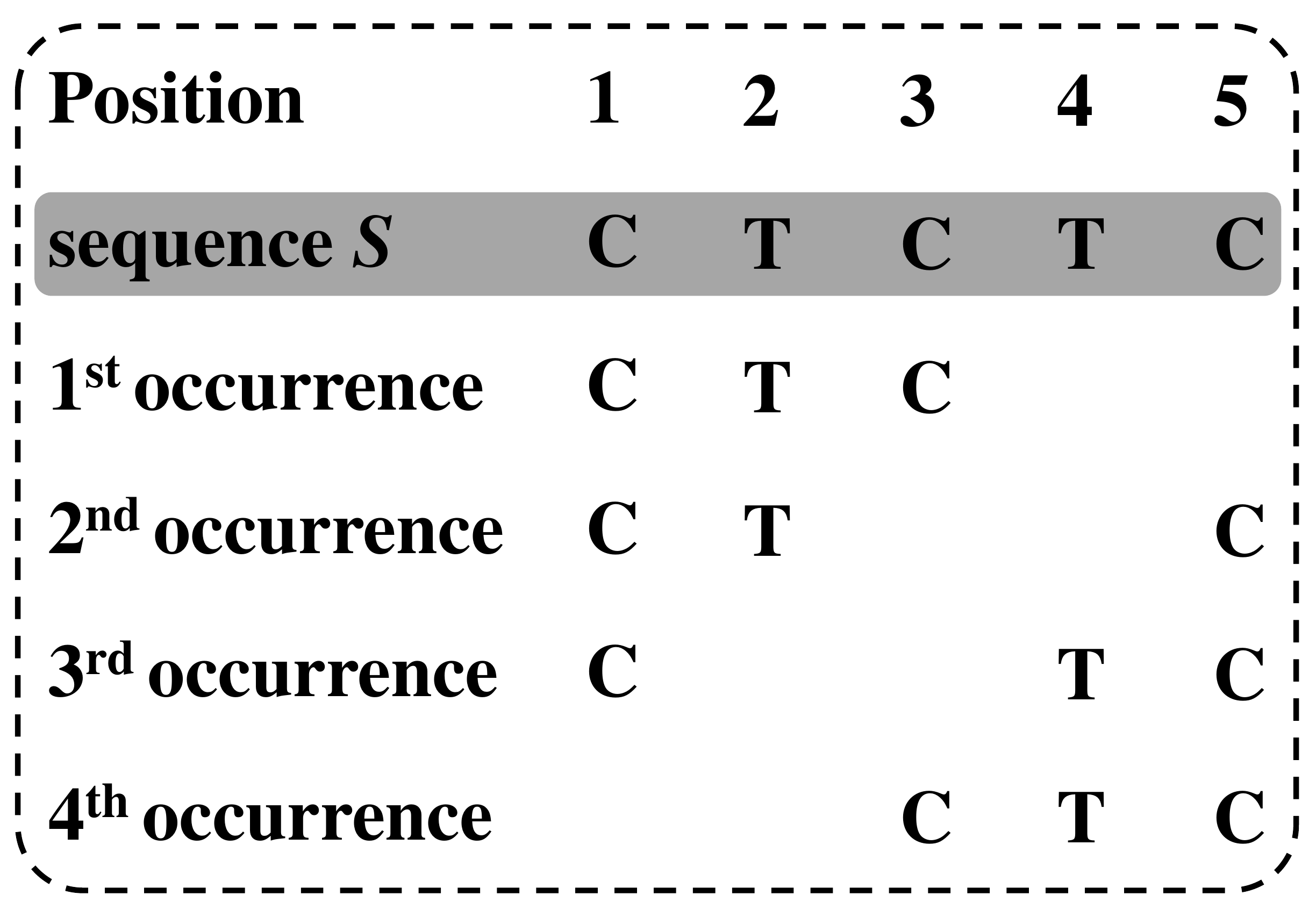}
    \caption{The occurrences of pattern $<$C, T, C$>$ in \textit{S}.}
    \label{fig:occurrence}
\end{figure}

The task of gap-constrained SPM has different types of conditions employed in various algorithms. The conditions are no-condition, one-off condition, and non-overlapping condition, respectively. Gap-constrained SPM with no-condition means that there are no conditions or restrictions on how  a letter appears in patterns. However, mining with no-condition does not meet the anti-monotone property (the Apriori property), i.e., the support of the super-patterns of a pattern is not larger than that of it. For example, given a sequence \textit{S} = "CTCTC", the gap constraint is $[$0, 2$]$, and the length constraint is ignored here. With no-condition, the support of the pattern $<$C, T$>$ is 3 but the support of its super-pattern $<$C, T, C$>$ is 4, which is shown in Fig. \ref{fig:occurrence}. Apparently, the Apriori property is not suitable for this situation. There are some algorithms \cite{wu2010nettree, guo2013pattern} based on the one-off condition of gap-constrained SPM, which means that it only allows a letter in sequence to be used once to match a letter in a pattern. For instance, in the preceding example, the support of $<$C, T$>$ is equal to 2, but the support of $<$C, T, C$>$ is reduced to 1. In Fig. \ref{fig:occurrence}, after the $\rm 1^{st}$ occurrence is found, the remaining three occurrences can no longer be found due to the one-off condition. The one-off condition is suitable for the Apriori property, but it may miss a lot of meaningful patterns. As a compromise between the above two mining conditions, non-overlapping \cite{wu2017nosep, wu2013pmbc} is widely used in gap-constrained SPM. When calculating the support of a pattern, if a letter appears more than once in this pattern, it cannot be counted twice at the same position in the sequence but can be counted at different positions. When calculating the support of $<$C, T, C$>$, it still considers the above example calculation. Due to the non-overlapping condition, after finding the $\rm 1^{st}$ occurrence, the occurrences of $\rm 2^{nd}$ and $\rm 3^{rd}$ can not be found because the first `C' is repeated in position 1 in $\rm 1^{st}$, $\rm 2^{nd}$, and $\rm 3^{rd}$ occurrences, but the $\rm 4^{th}$ occurrence can still be found because their repeated letters in position 3 represent the first `C' and second `C' in the pattern, respectively. Thus, it is more reasonable and relatively simpler to use the non-overlapping conditions to mine these patterns in bioinformatics sequence databases.

In response to the above requirements and challenges, and inspired by some previous works \cite{huang2022taspm, wu2017nosep}, we propose a novel and efficient algorithm named TALENT (\underline{TA}rgeted mining of Sequentia\underline{L} Patt\underline{E}r\underline{N} with Cons\underline{T}raints). It can be used not only to discover non-overlapping sequential patterns but also to discover desired patterns that contain query sequences input by users. Non-overlapping sequential pattern mining is highly applicable to biological data analysis. Therefore, mining biological data by the method of query sequence mining would have good application value in the targeted search for some disease-related DNA fragments or peptide chain molecules. For example, CpG islands\footnote{https://en.wikipedia.org/wiki/CpG\_island\_hypermethylation} is a region of DNA that contains a large number of linked cytosines (C), guanines (G), and phosphate bonds that connect the two above. CpG islands are often studied by the medical community for their methylation status \cite{liu2020role}. If users first obtain all DNA sequences and then select the DNA fragments of CpG islands, it is time-consuming and tedious. Such an inefficient method will lead to unnecessary pattern extensions. However, if users set the query sequence to a pattern formed by the combination of `C', sequences of phosphate bonds, and `G', a lot of invalid patterns can be filtered out. The preceding example shows that the proposed TALENT has an application value in biological data. In order to cope with specific processed data provided by different application scenarios, we propose two versions of the TALENT algorithm: \underline{B}readth-\underline{F}irst \underline{S}earch (BFS) and \underline{D}epth-\underline{F}irst \underline{S}earch (DFS) versions. In addition, we introduce a post-processing technique. These two approaches are designed from different traversal perspectives. More importantly, we develop four pruning strategies: \underline{S}equences \underline{P}re-\underline{R}ead \underline{P}runing strategy (SPRP strategy), \underline{I}tems \underline{P}re-\underline{R}ead \underline{P}runing strategy (IPRP strategy), \underline{B}readth-first \underline{P}re-\underline{E}xtend \underline{P}runing strategy (BPEP strategy, used in breadth-first search version of TALENT), and \underline{D}epth-first \underline{P}re-\underline{E}xtend \underline{P}runing strategy (DPEP strategy, used in depth-first search version of TALENT) to reduce the calculation of the algorithm from two aspects: sequence reading and pattern extension. The major contributions of this paper are as follows:

\begin{itemize}
    \item Targeted mining of non-overlapping sequential patterns is more meaningful and valuable than mining all patterns without a target set. There is no algorithm proposed to discover frequent target patterns with non-overlapping information. To address the shortcomings of current mining methods, we are the first to formalize the problem of targeted mining non-overlapping sequential patterns.

    \item For selecting the frequent target patterns that meet the conditions of \textit{minsup} and contain the query sequence set by users, we use a post-processing technique to filter these patterns. In addition, pruning strategies including SPRP and IPRP are used to reduce the number of sequences in the database and items in sequence, respectively.

    \item Although the post-processing technique can be conducted after all frequent patterns are generated, it takes a lot of unnecessary running time and memory. Thus, we further introduce two pruning strategies, called BPEP and DPEP, for the two versions of our algorithms to reduce the generation of redundant patterns.

    \item The empirical experiments on several real datasets show that using different strategies, the breadth-first search (BFS-version) and depth-first search (DFS-version) of TALENT improve to varying degrees over the baseline NOSEP$\rm _{Ta}$.
\end{itemize}

The following is the rest of this paper: In Section \ref{sec:relatedwork}, previous studies on SPM, non-overlapping SPM, and targeted pattern mining are briefly generalized. In Section \ref{sec:preliminaries}, the definitions of targeted mining of non-overlapping sequential patterns are introduced. Then, in Section \ref{sec:algorithm}, our efficient algorithm namely TALENT is proposed. In Section \ref{sec:experiments}, experimental evaluations are presented. In Section \ref{sec:conclusion}, we draw the conclusions and discuss the future work of this paper.
\section{Related Work} \label{sec:relatedwork}

\subsection{Sequential Pattern Mining} \label{section:SPM}

Sequential pattern mining (SPM) has become a popular technique since Agrawal and Srikant \cite{agrawal1995mining} proposed the Apriori property in 1995. As a groundbreaking algorithm, the Apriori property aims to search the items that occur frequently in the database, and the minimum support threshold, which is called \textit{minsup}, can be set by the user. It is the prototype of the earliest SPM. SPM is popular because users can discover patterns hidden in large databases that can be interpreted by humans \cite{fournier2017survey,huynh2022efficient,van2018mining}, which helps people get more information about the data and make better decisions. Since then, more and more algorithms for SPM have been designed to improve efficiency. For example, Srikant \textit{et al.} \cite{srikant1996mining} proposed the GSP algorithm to discover the generalized sequential patterns. Zaki \textit{et al.} \cite{zaki2001spade} proposed the SPADE algorithm to break the original problem into smaller sub-problems to improve efficiency. A prefix-projection method in PrefixSpan \cite{han2001prefixspan} can greatly reduce the size of the projection database. Besides, Ayres \textit{et al.} \cite{ayres2002sequential} proposed the SPAM algorithm to combine a vertical bitmap representation of the database with efficient support counting. However, this algorithm generates many candidates for handling dense datasets because of the combinatorial explosion. The LAPIN algorithm \cite{yang2007lapin} was developed to solve this problem straightforwardly. In order to quickly calculate the supports of sequential patterns, Salvemini \textit{et al.} \cite{salvemini2011fast} proposed the FAST algorithm, representing the datasets with indexed sparse id-lists. Fournier-Viger \textit{et al.} \cite{fournier2014fast} proposed the CM-SPAM algorithm with a new data structure named co-occurrence MAP (CMAP) for storing co-occurrence information. Gradually, various algorithms were designed according to different mining requirements. For example, top-$k$ SPM \cite{zhang2021tkus, chen2023towards} solves the problem that the minimum support threshold (\textit{minsup}) is difficult to set; utility-driven SPM \cite{gan2021survey,gan2019proum,gan2021fast} is used to search for some patterns that maybe infrequent but valuable; negative SPM \cite{cao2016nsp} can discover itemsets that do not appear but could be useful; constraint-based SPM \cite{bonchi2004closed} is designed to mine closed patterns. Up to now, there have been many study algorithms for SPM because sequential patterns are widely used in various research fields \cite{fournier2017survey}, such as bio-informatics \cite{wu2017nosep}, e-learning \cite{tarus2017hybrid}, text analysis \cite{zhong2010effective}, and web analysis \cite{tarus2018hybrid}.

\subsection{Non-overlapping Sequential Pattern Mining}
\label{section:NOSPM}

In the task of SPM, one of the crucial application fields is bioinformatics. However, traditional SPM is not suitable for mining patterns containing biological information in databases. It is meaningless to append the items that are far away from the mined sequence in a database to construct these sub-sequences. Contiguous patterns are more suitable for biological information mining. However, some site mutations may need to be considered when mining these contiguous patterns in bioinformatics. As a result, mining for motifs is proposed to mine contiguous patterns while accounting for noise tolerance \cite{floratou2011efficient}. There is a follow-up method that is intended to make the task of mining patterns in bioinformatics more flexible. This method allows users to set the constraints for gap and length instead of mining contiguous sequential patterns. The task of gap-constrained SPM employs three types of conditions in the various algorithms: no condition, one-off condition, and non-overlapping condition, in that order. Gap-constrained SPM without conditions is first studied for mining gap-constrained sequential patterns. However, mining without conditions does not meet the Apriori property. In order to solve this problem, Wu \textit{et al.} \cite{wu2013pmbc} proposed the PMBC algorithm based on the one-off condition of gap-constrained SPM. This means that it only allows an object in the sequence to be used once to match an object in a pattern. The one-off condition is suitable for the Apriori property, but may miss a lot of meaningful patterns. The third condition is a non-overlapping condition, which can mine as many patterns as possible, satisfying the Apriori property. When the non-overlapping condition is first proposed, the value of support can only be approximately calculated. Then, Wu \textit{et al.} \cite{wu2017nosep} created an innovative data structure called Nettree and put forward the first precise algorithm to discover non-overlapping sequential patterns efficiently. So far, the algorithms based on Nettree are the most efficient for mining non-overlapping sequential patterns. Recently, in order to reduce the difficulty of setting support, Chen \textit{et al.} \cite{chen2023towards} proposed the top-$k$ algorithms based on Nettree to discover $k$ non-overlapping sequential patterns with the highest support.

\subsection{Targeted Pattern Querying}
\label{section:TPM}

In some cases, the task of mining requires the inclusion of a few specified items, which are not necessary to generate all patterns. In order to improve the efficiency of mining, one of the most effective methods is to make targeted queries to reduce the number of redundant patterns. Association analysis techniques are the first application of targeted queries. In 2003, Kubat \textit{et al.} \cite{kubat2003itemset} proposed targeted association querying that uses a data structure called Itemset-Tree to mine rules that contain the itemset the user specified. In their work, they discuss three query cases, one of which is to perform the query by having users enter a target item. MEIT was proposed by Fournier-Viger \textit{et al.} \cite{fournier2013meit} to reduce unnecessary nodes in the Itemset-Tree in order to improve the efficiency of targeted queries in association mining. In addition, to solve the problem that Itemset-Tree does not meet the Apriori property and requires many invalid operations, Lewis \textit{et al.} \cite{lewis2019enhancing} proposed a generation process to generate rule sets faster. Recently, Miao \textit{et al.} \cite{miao2021targetum} proposed the first algorithm for mining target high-utility itemsets.

Since sequence data differs greatly from transaction data, Chiang \textit{et al.} \cite{chiang2003goal} proposed an algorithm for mining goal-oriented sequential patterns. However, this approach only supports goal queries on the last itemset, which is limited in many application scenarios. For gap-constrained SPM, Chueh \textit{et al.} \cite{chueh2010mining} designed an algorithm to find target-oriented sequential patterns with time interval constraints. Chand \textit{et al.} \cite{chand2012target} utilized the constraints of recency and monetary to discover target-oriented sequential patterns. However, these two algorithms have the same limitations. In order to solve this problem, Huang \textit{et al.} \cite{huang2022taspm} proposed the TaSPM algorithm to discover target sequential patterns. Distinct from the previous algorithms, the consideration of the mining problem is more complex and comprehensive. Inspired by previous works like TaSPM, we concentrate on the task of mining target sequential patterns with non-overlapping information and propose a novel and efficient algorithm to better solve the gap-constrained tasks in bioinformatics.
\section{Preliminaries and Problem Statement}
\label{sec:preliminaries}

In this section, we give some basic concepts and notations about frequent target sequential patterns and non-overlapping sequential patterns. And the problem definition of targeted mining of non-overlapping sequential patterns is formalized. In order to allow readers to have a better understanding of this work and its connection to previous studies, some related definitions are adopted from previous research works. The related definitions are given as follows.

\begin{definition}[Sequence database]	
    \label{definition 1}
    \rm In a database, there are a series of items, stored in a finite set $I$ = \{$i_{1}$, $i_{2}$, $i_{3}$, ..., $i_{N}$\}. For example, in DNA sequences, the series of items are \{A, C, G, T\} with $N$ = 4. $X$ is an itemset that is a subset of $I$. The itemset $X$ in a sequential pattern may contain one or more items. However, in the task of non-overlapping sequential pattern mining, an itemset usually has only one item for the sake of convenience \cite{wu2017nosep}. In other words, the lengths of all these itemsets are 1, represented by 1-itemsets. Similarly, a pattern with a length of $\xi$ is a $\xi$-pattern, where $\xi$ is also the size of itemsets in this pattern. In addition, for a sequence $S$ = $<$$X_1$, $X_2$, $X_3$, ..., $X_n$$>$, the size of $S$ is $n$. For our mining tasks, the length of $S$ is the same as its size. The items in the itemset occur in order, i.e., $X_{i-1}$ occurs before $X_{i}$ in order. If the positions of $X_{i-1}$ and $X_{i}$ are changed, the original pattern is changed too. It is just like buying a piece of bread after buying a bar of chocolate in the supermarket, which is not the same as buying bread and then buying chocolate. Additionally, a sequence database is defined as $\mathcal{D}$ = \{$S_{1}$, $S_{2}$, $S_{3}$, ..., $S_{n}$\}. In every sequence in $\mathcal{D}$, there is a unique identifier \textit{SID} and the data sequence itself \textit{DS}, forming a tuple of the sequence database. 
\end{definition}

\begin{definition}[Support of pattern]
    \label{definition 2}
    \rm The support of a pattern $P$ in the sequence $S$ is denoted by \textit{sup}($P$, $S$), which indicates the frequency of the pattern $P$ in $S$. In our tasks, it is usually expressed as the number of occurrences. In the whole sequence database $\mathcal{D}$, the support is represented by \textit{sup}($P$, $\mathcal{D}$) and can be abbreviated as \textit{sup}($P$). It is calculated as the sum of supports of all sequences in the database. Mathematically, \textit{sup}($P$) is equal to \(\sum_{i = 1}^{n}|\textit{sup}(P,S_{i})|\), where $n$ is the number of sequences in the database. In algorithms for sequential patterns, the minimum support threshold is usually denoted as \textit{minsup}, and it is set by the user. Sequential pattern mining (SPM) seeks all sequential patterns with supports greater than or equal to \textit{minsup}.
\end{definition}

\begin{table}[h]
	\centering
	\caption{Sequence database $\mathcal{D}$}
	\label{table1}
	\begin{tabular}{|c|c|}  
		\hline 
		\rm \textbf{SID} & \rm \textbf{DS} \\
		\hline  
		\(S_{1}\) & \rm A T C A C T C G \\ 
		\hline
		\(S_{2}\) & \rm T G G C T \\ 
		\hline
		\(S_{3}\) & \rm A G T A A \\  
		\hline  
		\(S_{4}\) & \rm G A G A T G \\  
		\hline  
	\end{tabular}
\end{table}

As shown in Table \ref{table1}, \textit{sup}($<$A$>$, $S_{1}$) = 2, \textit{sup}($<$A$>$, $S_{2}$) = 0, \textit{sup}($<$A$>$, $S_{3}$) = 3,  and \textit{sup}($<$A$>$, $S_{4}$) = 2, so \textit{sup}($<$A$>$) = 2 + 0 + 3 + 2 = 7. Similarly, \textit{sup}($<$C$>$) = 4, \textit{sup}($<$G$>$) = 7, and \textit{sup}($<$T$>$) = 6. If \textit{minsup} is set to 5, then pattern $<$C$>$ is not considered as a frequent pattern.

\begin{definition}[Non-overlapping conditions and gap constraints \rm \cite{wu2017nosep}]
    \label{definition 3}
    \rm In Section \ref{sec:relatedwork}, some prior knowledge such as non-overlapping conditions has been stated. The mined pattern $P$ with constraints is denoted as $p_{1}$ $\left[min_{1}, max_{1}\right]$ $p_{2}$ ... $\left[min_{m-1}, max_{m-1}\right]$ $p_{m}$. In our tasks, for convenience, denoting \textit{gap} = [\textit{mingap}, \textit{maxgap}], where \textit{mingap} = $min_{1}$ = $min_{2}$ = $min_{3}$ = ... = $min_{m-1}$ and \textit{maxgap} = $max_{1}$ = $max_{2}$ = $max_{3}$ = ... = $max_{m-1}$. The value of \textit{gap} can be set to a suitable pair of non-negative integers according to users' requirements in a mining task. In addition, for the sake of convenience, the occurrences are used to represent the position of an item in the sequence, i.e., the occurrences of a pattern are represented as <$O_1$, $O_2$, ..., $O_s$>, where $O_k$ is the occurrence position of the $k^{th}$ item of the pattern in the sequence and $s$ is the size of the pattern. When calculating the support of a pattern, if a letter $\mathcal{L}$ appears more than once in this pattern, then $\mathcal{L}$ cannot be counted twice at the same position of this pattern in the sequence but can be counted at different positions. In other words, only when two occurrences of a pattern subject to $O_k$ are all different in occurrences, then they need to be both counted when calculating the support under the non-overlapping condition, where $k$ = \{1, 2, ..., $s$\}.
\end{definition}

For instance in $S_4$, if \textit{gap} is set to $[$0, 2$]$ and \textit{len} is set to [1, 6], the occurrences of $<$G, A, G$>$ are $<$1, 2, 3$>$, $<$1, 4, 6$>$ and $<$3, 4, 6$>$. The occurrence of $<$1, 2, 6$>$ does not meet the gap constraints. Due to the non-overlapping condition, the occurrences $<$1, 2, 3$>$ and $<$1, 4, 6$>$ can not be counted twice when calculating the support, because $O_1$ of two occurrences are both 1. Instead, the occurrences $<$1, 2, 3$>$ and $<$3, 4, 6$>$ are both counted under the non-overlapping condition for the reason that $O_k$ are all different in occurrences $<$1, 2, 3$>$ and $<$3, 4, 6$>$, where $k$ = \{1, 2, 3\}. Hence, \textit{sup}($<$G, A, G$>$) is 2.

For the sequence database $\mathcal{D}$ shown in Table \ref{table1}, if \textit{gap} is set to $[$0, 2$]$, \textit{sup}($<$A, G$>$, $S_{1}$) = 0 because the occurrences of $<$1, 8$>$ and $<$4, 8$>$ are both fail to meet the gap constraints. By a simple calculation, \textit{sup}($<$A, G$>$, $S_{2}$) = 0, \textit{sup}($<$A, G$>$, $S_{3}$) = 1 and \textit{sup}($<$A, G$>$, $S_{4}$) = 2. Hence, \textit{sup}($<$A, G$>$) = 0 + 1 + 0 + 2 = 3. When calculating the \textit{sup}($<$G, A$>$, $S_{4}$), there are three occurrences which are $<$1, 2$>$, $<$1, 4$>$, $<$3, 4$>$ matches the pattern, but the letter in first positions of $<$1, 2$>$ and $<$1, 4$>$ is overlapping. Therefore, \textit{sup}($<$G, A$>$, $S_{4}$) equals to 2 rather than 3.

\begin{definition}[Length constraints \rm \cite{wu2017nosep}]
    \label{definition 4}
    \rm The length constraints are the condition of constraints of span length, which is the distance between the most anterior position ($p_A$) and the last position ($p_L$) of occurrences. Therefore, the span length can be calculated by $p_L - p_A + 1$. The length constraints are represented as \textit{len} = [\textit{minlen}, \textit{maxlen}], which means that when mining sequences, the span length can not be less than \textit{minlen} and can not be greater than \textit{maxlen}. If conditions are not met when mining, it is not counted in the support calculation.
\end{definition}

For example, in the sequence database $\mathcal{D}$, let \textit{gap} = [0, 3] and \textit{len} = [1, 6]. Given the pattern $<$A, A, G$>$, \textit{sup}($<$A, A, G$>$, $S_{2}$) = \textit{sup}($<$A, A, G$>$, $S_{3}$) = 0. In $S_{1}$, the occurrence of $<$A, A, G$>$ is $<$1, 4, 8$>$, which meets the gap constraints. However, since the corresponding $p_A$ = 1 and $p_L$ = 8, the span length is 8 - 1 + 1 = 8, which does not satisfy the length constraints. Thus, \textit{sup}($<$A, A, G$>$, $S_{1}$) = 0. While the occurrence of $<$A, A, G$>$ is $<$2, 4, 6$>$ in $S_{4}$ and it meets both the gap and length constraints. Therefore, \textit{sup}($<$A, A, G$>$, $S_{1}$) = 1 and \textit{sup}($<$A, A, G$>$) = $\sum_{i = 1}^{4}$ \textit{sup}($<$A, A, G$>$, $S_{i}$) = 1. 

\begin{definition}[Frequent target pattern and targeted pattern mining \rm \cite{huang2022taspm, miao2021targetum}]
    \label{definition 5}
    \rm Given a query sequence denoted as \textit{qs}, is used as the query condition set by the users according to their need for a targeted pattern mining task. A frequent target sequential pattern $P$ must satisfy \textit{sup}($P$) $\ge$ \textit{minsup} and \textit{qs} $\subseteq P$. In order words, the frequent target pattern $P$ must include \textit{qs}, and it also has enough support value. It should be noted that the frequent patterns in this task are different from the previous descriptions in traditional SPM. Obviously, targeted pattern mining aims to complete the task of mining all the frequent target patterns in a database. Targeted non-overlapping sequential pattern mining is aimed at discovering all the frequent target patterns with the non-overlapping condition and sequential constraints.
\end{definition}

For example, given the \textit{qs} = $<$A, T$>$, \textit{minsup} = 2,  \textit{gap} =$[$0, 2$]$ and \textit{len} = $[$1, 6$]$. In the database $\mathcal{D}$ shown in Table \ref{table1}, the support of $P_{1}$ = $<$A, T$>$ is 4 $\ge$ \textit{minsup} and \textit{qs} is included in $P_{1}$. Hence, it is regarded as a frequent pattern. And $P_{2}$ = $<$A, T, A$>$ is also a frequent target pattern because the support of $P_{2}$ is 2 $\ge$ \textit{minsup} and $P_{2}$ is a super-pattern of \textit{qs}. However, since the support of $P_{3}$ = $<$A, T, A, A$>$ is 1 \textless \textit{minsup}. Therefore, it is infrequent. In addition, although the support of $P_{4}$ = $<$A, A$>$ is 4 $\ge$ \textit{minsup}, \textit{qs} $\nsubseteq$ $P_{4}$. Therefore, $P_{4}$ is not regarded as a frequent target pattern as well.

\textbf{Problem statement:} Given a sequence database $\mathcal{D}$, the minimum support threshold \textit{minsup}, a query sequence \textit{qs}, gap constraints \textit{gap}, and length constraints \textit{len}. The task of targeted non-overlapping sequential pattern mining is to search all the patterns $P$ which meet the condition of the gap and length constraints, and they also satisfy that \textit{sup}($P$) $\ge$ \textit{minsup} and \textit{qs} $\subseteq$ $P$.
\section{The TALENT Algorithm} \label{sec:algorithm}

Based on the discussions and definitions mentioned above, this paper aims to show a novel and effective algorithm named TALENT towards target non-overlapping sequential patterns. The calculation of support and the generation of patterns are the most important operations of the mining tasks. We design powerful pruning strategies to boost efficiency. This section can be subdivided into three subsections to clarify the algorithm in detail, as follows. In Section \ref{NETGAP algorithm}, the algorithm based on the structure of Nettree to calculate the support is introduced. Subsequently, we propose the \underline{B}readth-\underline{F}irst \underline{S}earch (BFS) version and \underline{D}epth-\underline{F}irst \underline{S}earch (DFS) version of TALENT in Section \ref{TALENT_BFS} and Section \ref{TALENT_DFS} to solve problems of pattern generation, respectively.

\subsection{The Nettree-based Algorithm}
\label{NETGAP algorithm}

When calculating the support of the non-overlapping sequential patterns, the occurrences are difficult to search without using the backtracking strategy. For instance, given the sequence $S_{2}$ = "TGGCT" in Table \ref{table1} when $P_{1}$ = $<$T, G, T$>$ with \textit{gap} = [0, 1], the process of generating patterns are described in detail in Section \ref{TALENT_BFS} and Section \ref{TALENT_DFS}. In this section, we give a brief introduction. Firstly, the support of 1-pattern $<$T$>$ is calculated. They are easily calculated by a simple traversal. Secondly, the support of 2-pattern $<$T, G$>$ is calculated. The occurrences of $<$T, G$>$ are $<$1, 2$>$ and $<$1, 3$>$. However, due to the non-overlapping condition, the occurrence of $<$1, 3$>$ is not to be counted repeatedly. Finally, if the support of 3-pattern $<$T, G, T$>$ is calculated, the occurrence $<$1, 2, 5$>$ does not satisfy the gap constraints. Hence, in some algorithms, if the occurrence $<$1, 3, 5$>$ is calculated according to the occurrence of its sub-pattern $<$1, 3$>$, then the occurrence $<$1, 3, 5$>$ is possible to be an omission without a backtracking strategy, since $<$1, 3$>$ is not recorded. The \textit{NETGAP} algorithm based on the Nettree data structure proposed by Wu \textit{et al.} \cite{wu2017nosep} is used to solve this problem. It will be adopted in the following context to calculate the support. We give a definition and explanation of the Nettree structure and the \textit{NETGAP} algorithm where the query sequence \textit{qs} is ignored.

\begin{table}[h]
	\centering
	\caption{Pattern $P$ and sequence $S$}
	\label{table2}
	\begin{tabular}{|c|cccccccccccccccc|}
		\hline
		\rm \textbf{Pattern} & \multicolumn{4}{c}{$p_{1}$} & \multicolumn{4}{c}{$p_{2}$} & \multicolumn{4}{c}{$p_{3}$} & \multicolumn{4}{c|}{$p_{4}$} \\
		\hline
		$P$ & \multicolumn{4}{c}{\rm T} & \multicolumn{4}{c}{\rm C} & \multicolumn{4}{c}{\rm A} & \multicolumn{4}{c|}{\rm G} \\
		\hline
		\rm \textbf{Sequence} & $s_{1}$ & $s_{2}$ & $s_{3}$ & $s_{4}$ & $s_{5}$ & $s_{6}$ & $s_{7}$ & $s_{8}$ & $s_{9}$ & $s_{10}$ & $s_{11}$ & $s_{12}$ & $s_{13}$ & $s_{14}$ & $s_{15}$ & $s_{16}$\\
		\hline
		$S$ & \rm G & \rm T & \rm C & \rm A & \rm A & \rm G & \rm T & \rm C & \rm T & \rm C & \rm T & \rm C & \rm A & \rm G & \rm G & \rm T \\
		\hline
	\end{tabular}
\end{table}

\begin{figure}[ht]
    \centering
    \includegraphics[clip,scale=0.2]{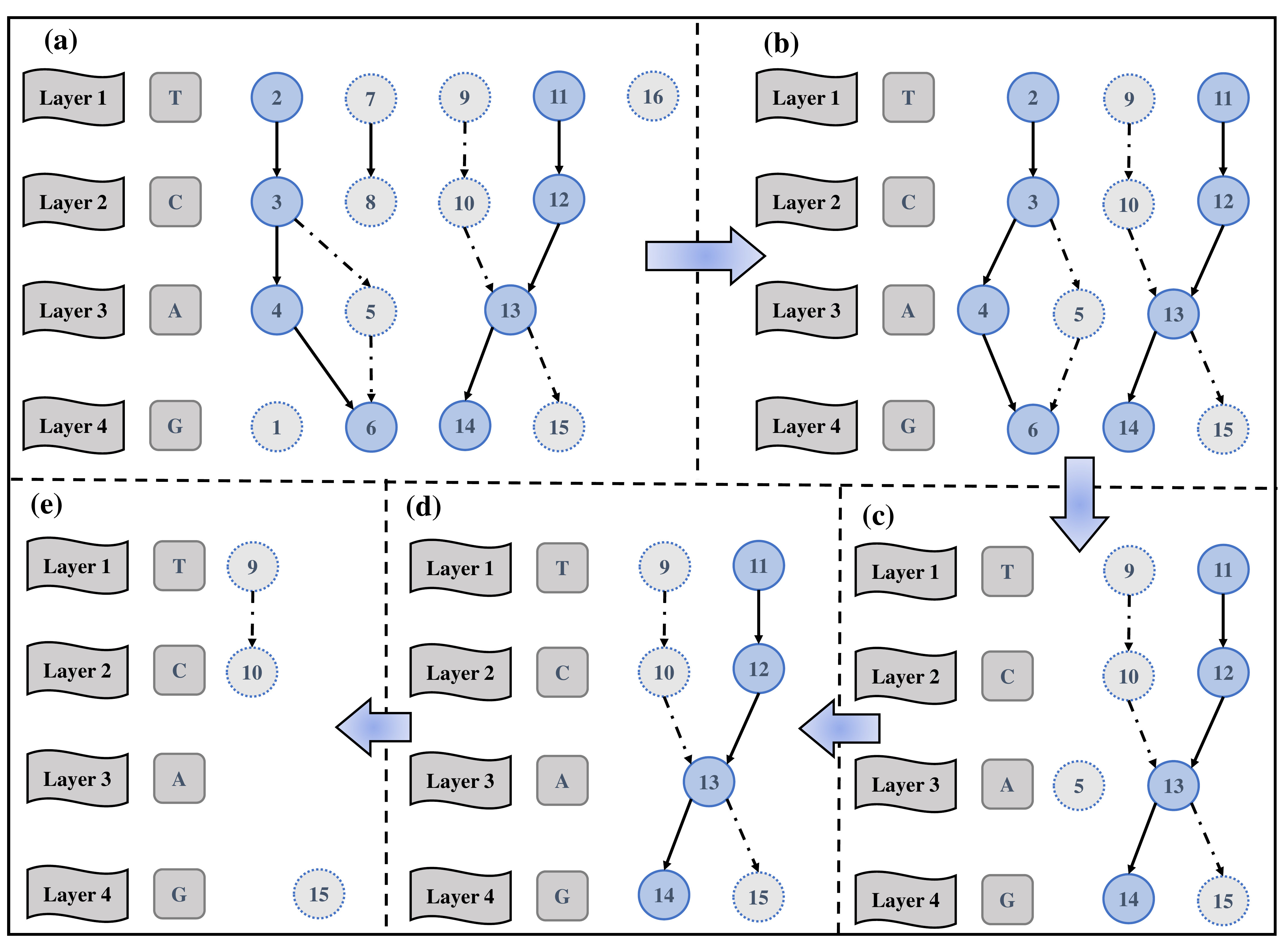}
    \caption{The Nettree for the pattern $<$T, C, A, G$>$ in the sequence $S$.}
    \label{Nettree}
\end{figure}

\begin{definition}[Nettree structure \rm \cite{wu2017nosep, wu2010nettree}]
    \label{definition 6}
    \rm As the name implies, Nettree is a tree-like structure with the property of a net. Similar to the tree, Nettree has elements of \textit{parent}, \textit{child}, \textit{level}, \textit{root}, and \textit{leaf}. However, the Nettree has some different properties from the tree. There is only one \textit{root} in a tree, but maybe more than one \textit{roots} in a Nettree. In order to easily represent the nodes in a Nettree, we define the node $i$ in the $j^{th}$ \textit{level} as $n^{i}_{j}$. In addition, it is known that all the non-\textit{root} nodes of the tree have one parent. Another feature of Nettree differs from the tree is that every non-\textit{root} node of the Nettree may have one or more parents, but there is a limitation that all its \textit{parents} must be in the same \textit{level}. Thus, differing from the uniqueness of the path from the non-\textit{root} node to \textit{root} node of the tree, a non-\textit{root} node of Nettree has multiple paths to \textit{root} node. Consequently, we give one definition that the \textit{leaves} in the last \textit{level} of Nettree are defined as absolute \textit{leaves} and a full path represents a path from a \textit{root} node to an absolute \textit{leaf} node.
\end{definition}

As shown in Table \ref{table2}, given a sequence $S$ = \{$s_{1}s_{2}s_{3}s_{4}s_{5}s_{6}s_{7}s_{8}s_{9}s_{10}s_{11}s_{12}s_{13}s_{14}s_{15}s_{16}$\} = "GTCAAGTCT\\CTCAGGT". The gap constraint is set to \textit{gap} = [0, 2], and the length constraint is set to \textit{len} = [1, 5]. Whenever a pattern is generated, the \textit{NETGAP} algorithm creates a new Nettree to calculate the support. When mining the pattern $P$ = $<$T, C, A, G$>$, \textit{NETGAP} creates the Nettree as shown in Fig. \ref{Nettree}. And the algorithm is shown in Algorithm \ref{alg:NETGAP}.

Actually, the goal of \textit{NETGAP} is to create and update the Nettree to calculate the support of a pattern $P$ in sequence $S$, which is denoted as \textit{sup}($P$, $S$). In addition, the support of a pattern in a database $\mathcal(D)$ of multiple sequences which are represented as \textit{sup}($P$, $\mathcal{D}$) is calculated by  \textit{sup}($P$, $\mathcal{D}$) = $\sum_{1}^{n}$ \textit{sup}($P$, $S_{i}$), where $n$ is the number of sequences in the database. In order to elaborate the \textit{NETGAP} in detail, we divide it into four parts as follows:

\begin{itemize}
    \item \textbf{STEP 1:} First of all, at level $i$, we list the positions in sequence $S$, that correspond to the $i$-th element of $P$ as the leave nodes in level $i$. For example, at level 1, we have leave nodes corresponding to the positions, 2, 7, 9, 11, and 16 of $S$ that contain $T$. Next, the \textit{leaf} nodes of two adjacent \textit{levels} should be determined if they need to be connected. Given two nodes $n^{i_1}_{j_1}$ and $n^{i_2}_{j_2}$ where $i_1$ \textless $ i_2$ and $j_1$ \textless $j_2$, and if $j_2$ = $j_1$ + 1 and \textit{mingap}  $\le i_2 - i_1 + 1 \le$  \textit{maxgap}, then connect the nodes $n^{i_1}_{j_1}$ and $n^{i_2}_{j_2}$. In brief, two nodes in Nettree which are in adjacent \textit{levels} and meet the gap constraints should be connected. For instance, in Fig. \ref{Nettree} (a), $n_{1}^{2}$ and $n_{2}^{3}$ need to be connected. Similarly, $n_{2}^{3}$ should be connected with $n_{3}^{4}$ and $n_{3}^{4}$ should be connected with $n_{4}^{6}$.

    \item \textbf{STEP 2:} Secondly, the \textit{leaf} nodes with no path to reach the nodes of previous \textit{level} are called lonely nodes, such as $n_{4}^{1}$ in Fig. \ref{Nettree} (a). Similarly, the nodes which are not the absolute \textit{leaf} with no path to reach the nodes of the next \textit{level} are lonely nodes as well, such as $n_{1}^{16}$. It can be obtained by an easy derivation that the lonely nodes can be pruned that the calculation of support will not be affected. This process requires repeating pruning. For example, in Fig. \ref{Nettree} (a), the $n_{4}^{1}$, $n_{2}^{8}$ and $n_{1}^{16}$ are lonely nodes which should be pruned. After pruning the node $n_{2}^{8}$, the node $n_{1}^{7}$ become a lonely node, too. Thus, it should be pruned as well. Hence, after the step, the Nettree is pruned as shown in Fig. \ref{Nettree} (b).

    \item \textbf{STEP 3:} Thirdly, the support of $P$ in $S$ is calculated. Initially, the support is assigned a value of 0. Apparently, every full path means that a complete match of $P$ in $S$ such as $<$2, 3, 4, 6$>$ in Fig. \ref{Nettree} (b). Hence, after traversing a full path, the support needs to be increased by 1. After updating the support, this path is pruned as shown in Fig. \ref{Nettree} (c). Then, the newly created single nodes are also pruned, as shown in Fig. \ref{Nettree} (d). It prevents the overlapping sequential patterns from being counted twice. And it is repeated until no full path exists, as shown in Fig. \ref{Nettree} (e). In particular, the span length of occurrence $<$9, 10, 13, 14$>$ is calculated by 14 - 9 + 1 = 6 \textgreater \textit{maxgap} = 5, which does not meet the length constraints. Hence, it is not considered a complete path.
\end{itemize}

\begin{algorithm}[h]
	\caption{The NETGAP algorithm}
	\label{alg:NETGAP}
	\LinesNumbered
	\KwIn{sequence $S$; gap constraint \textit{gap} = [\textit{mingap, maxgap}] and length constraint \textit{len} = [\textit{minlen, maxlen}]; the mined pattern $P$.}
	\KwOut{\textit{sup}($P$, $S$).}
	create the nodes of $P$ in $S$\;
	connect the nodes of occurrences of two adjacent \textit{levels} which satisfy the gap constraints; \qquad $//$ STEP 1 \\
	\While{\rm single nodes exist}{
		prune the nodes if they are lonely nodes; \qquad $//$ STEP 2 \\
	}
	\textit{sup}($P$, $S$) $\leftarrow$ 0\;
	\For{$n_{1}^{i} \in $ \textit{Nettree} \do}{
		\textit{node}[1] $\leftarrow$ $n_{1}^{i}$;\qquad $//$ occurrences are stored in nodes\\
		\For{\rm $j$ = 1 to \textit{Nettree.level} -1 \do}{
			\textit{node}[$j$+1] = the leftmost \textit{child} of \textit{node}[$j$] satisfied the length constraints\;
		}
		\textit{sup}($P$, $S$) $\leftarrow$ \textit{sup}($P$, $S$) + 1\;
		prune the \textit{node}[$j$]; \qquad $//$ STEP 3
	}
	\textbf{return} \textit{sup}($P, S$)
\end{algorithm}

\subsection{The Breadth-First Search Version of TALENT Algorithm}
\label{TALENT_BFS}

\begin{definition}[Prefix and suffix \rm \cite{wu2017nosep}]
    \label{definition 7}
    \rm The prefix of a pattern $P$ is the sub-pattern of the first $n$ items, where $n$ is the size of $P$ - 1. In the same way, the suffix of a pattern $P$ is the sub-pattern of the last $n$ items.
\end{definition}

For instance, given a pattern $P$ = $<$T, C, G, T, A$>$, the size of $P$ is 5 and $n$ is equal to 4. Hence, the prefix of $P$ is $<$T, C, G, T$>$ and the prefix is $<$C, G, T, A$>$.

\begin{algorithm}[h]
	\small
	\caption{The gen\_candidate\_BFS algorithm}
	\label{alg:candidate}
	\LinesNumbered
	\KwIn{\textit{cand}[\textit{len}].}
	\KwOut{new candidates set $S_C$}
	\textit{start} $\leftarrow$ 1\;
	\For{\rm $i$ = 1 to \textit{len}}{
		$Q \leftarrow$ \textit{preffix}(\textit{cand}[\textit{len}][\textit{start}])\;
		$R \leftarrow$ \textit{suffix}(\textit{cand}[\textit{len}][$i$])\;
		\If{!$Q$.EqualTo($R$)}{
			\textit{start} $\leftarrow$ binarySearch(\textit{cand}[\textit{len}], $R$, 1, \textit{len})\;
		}
		\If{\rm  \textit{start} $\ge$ 1 and \textit{start} $\le$ \textit{len}}{
			\While{\rm $Q$.equalTo($R$)}{
				$C_{con}$ = \textit{cand}[\textit{len}][\textit{i}].add(\textit{cand}[\textit{len}][\textit{start}.finalItem])\;
				$S_C$.append($C_{con}$)\;
				\textit{start}++\;
				\If{\textit{start} \textgreater \textit{len}}{
					\textit{start} $\leftarrow$ 1\;
					$Q \leftarrow$ prefix(\textit{cand}[\textit{len}][\textit{start}])\;
				}
			}
		}
	}
	\textbf{return} $S_C$
\end{algorithm}

\begin{figure}[ht]
    \centering
    \includegraphics[clip,scale=0.19]{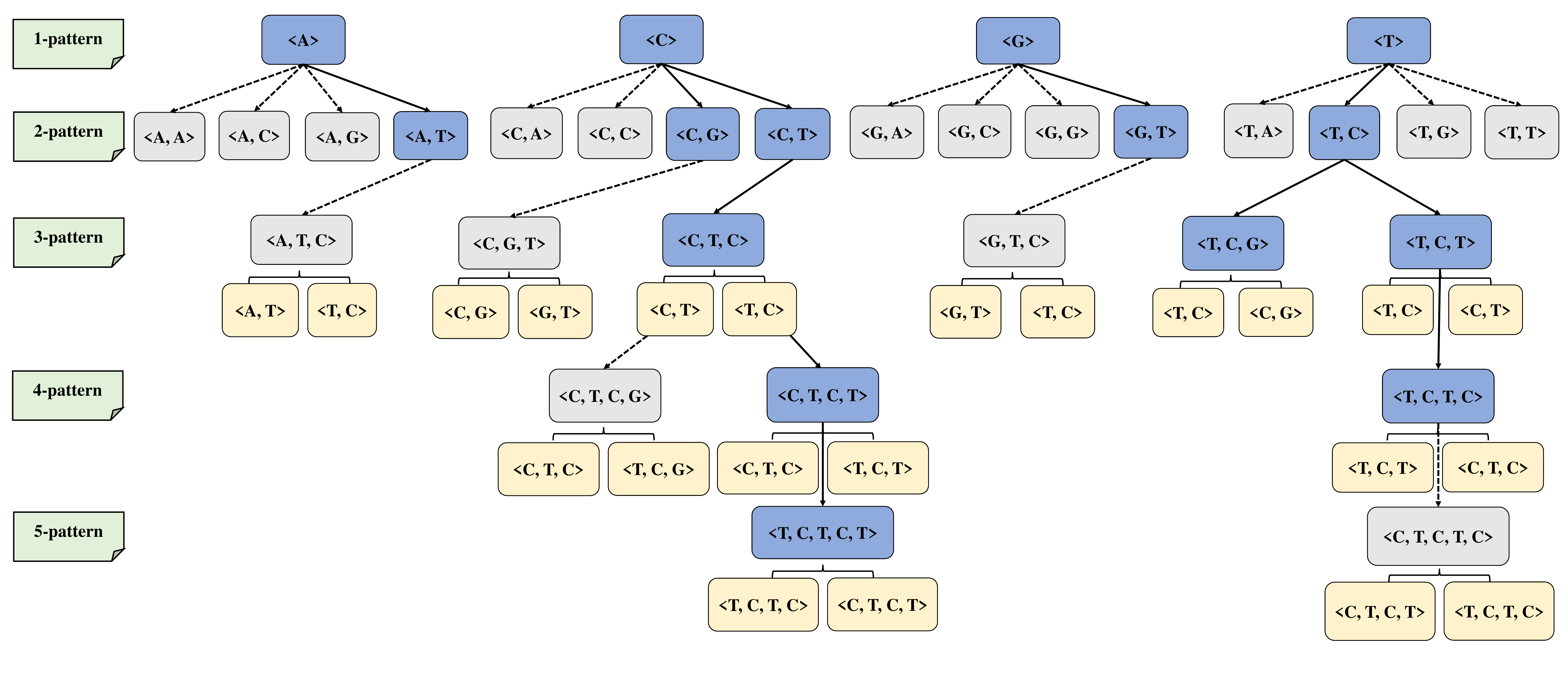}
    \caption{The process of pattern generation in BFS way.}
    \label{geneate_BFS}
\end{figure}

As mentioned above, in addition to computing support for mining non-overlapping sequential patterns, another important issue that needs to be addressed is the generation of patterns. Wu \textit{et al.} \cite{wu2017nosep} proposed an effective method named \textit{gen\_candidate} to generate the patterns. It is shown in Algorithm \ref{alg:candidate}. The main idea of this method is to achieve it through concatenation. To be specific, the $r$-patterns are generated by the concatenation of two $q$-patterns, where $r = q + 1$. For example, when mining non-overlapping sequential patterns in the sequence $S$ = "GTCAAGTCTCTCAGGT" in Table \ref{table2} with the condition of \textit{gap} = [0, 3], \textit{len} = [1, 8] and \textit{minsup} = 3 without query sequence \textit{qs}, the process is shown in Fig. \ref{geneate_BFS}. To begin with, the algorithm generates the 1-patterns which are $<$A$>$, $<$C$>$, $<$G$>$, and $<$T$>$. Because they are all frequent patterns, the 2-patterns are generated without pruning. With the calculation of the \textit{NETGAP} algorithm, the frequent 2-patterns are $<$A, T$>$, $<$C, G$>$, $<$C, T$>$, $<$G, T$>$ and $<$T, C$>$. When the size of the pattern is larger than 2, the patterns are generated through the patterns of the previous size, which are called candidates. If 2-patterns are to become 3-patterns by adding extensions directly from behind, 4 × 5 = 20 patterns will be generated. Hence, an effective method is proposed. For example, the pattern $<$A, C$>$ is not frequent, and thus the super-patterns of it such as $<$A, C, T$>$ and $<$T, A, C$>$ are bound to be infrequent. Based on this inference, the candidates of length 2 are generated the prefixes and the suffixes of 3-patterns. In the sequence $S$, $<$A, T$>$ and $<$T, C$>$ are both frequent. The 3-pattern $<$A, T, C$>$ that are concatenated from $<$A, T$>$ and $<$T, C$>$ is possible to be a frequent pattern. Therefore, the number of generated patterns will be significantly reduced. Among all 3-patterns, since only $<$A, T, C$>$, $<$C, G, T$>$, $<$C, T, C$>$, $<$G, T, C$>$, $<$T, C, G$>$, and $<$T, C, T$>$ are generated, the number is 6, which is far smaller than 20. In order to search the suffix which is corresponding to the prefix more efficiently, the method of binary search is used here. In this way, we perform the above steps until there is no generated pattern in a new layer or all generated patterns do not meet the support condition.

The algorithm of \textit{gen\_candidate} is suitable for breadth-first search (BFS) of non-overlapping sequential patterns. After introducing the calculation of the support degree and the generation of the pattern, the BFS version of TALENT is described in detail. In the following, we start with a definition of the post-processing technique.

\begin{definition}[Post-processing technique \rm \cite{huang2022taspm}]
    \label{definition 8}
    \rm As mentioned in Section \ref{sec:relatedwork}, there is still no algorithm towards the new definition of frequent target non-overlapping sequential patterns. Therefore, if there is a need to implement it by utilizing the traditional algorithm to mine the non-overlapping sequential patterns, a post-processing technique is needed. After the patterns whose support is larger than \textit{minsup} are mined, they need to be judged whether contain the query sequence \textit{qs}. Through this process, a pattern containing the query sequence will be saved. Otherwise, it will be removed. Because the process of judging whether to include \textit{qs} is after all the patterns that meet the condition of support are obtained, this process of judging is named a post-processing technique.
\end{definition}

For instance, let the non-overlapping sequential patterns mined in the sequence $S$ = "GTCAAG\\TCTCTCAGGT" in Table \ref{table2} with the condition of \textit{gap} = [0, 3], \textit{len} = [1, 10] and \textit{minsup} = 3 without query sequence \textit{qs}. After generating the patterns and calculating their support, there are 15 frequent sequential patterns, which are $<$A$>$, $<$C$>$, $<$G$>$, $<$T$>$, $<$A, T$>$, $<$C, G$>$, $<$C, T$>$, $<$G, T$>$, $<$T, C$>$, $<$C, T, C$>$, $<$T, C, G$>$, $<$T, C, T$>$, $<$C, T, C, T$>$, $<$T, C, T, C$>$, and $<$T, C, T, C, T$>$, respectively. If the query sequence \textit{qs} are set to $<$T, C$>$, the patterns that do not contain $<$T, C$>$ will be filtered through the post-processing technique. Hence, the frequent sequential patterns are $<$T, C$>$, $<$C, T, C$>$, $<$T, C, G$>$, $<$T, C, T$>$, $<$C, T, C, T$>$, $<$T, C, T, C$>$ and $<$T, C, T, C, T$>$, where the number is reduced to 7. The way that patterns generate is shown in Fig. \ref{geneate_BFS}. The patterns whose annotation color is gray with the support of less than \textit{minsup}. The super-patterns of these patterns will not be generated including the cases of being as a prefix as well as a suffix.

After the \textit{gen\_candidate} algorithm generates the new patterns, the \textit{NETGAP} algorithm calculates the support, and the post-processing technique selects out the patterns that meet the condition of containing the query sequence, the frequent target non-overlapping sequential patterns can be mined. However, this method is not efficient enough. In order to reduce the computational complexity, we propose several pruning strategies for the breadth-first search of non-overlapping sequential patterns.

\begin{table}[h]
	\centering
	\caption{Sequence database $\mathcal{D}$}
	\label{table3}
	\begin{tabular}{|c|c|}
		\hline
		\rm \textbf{SID} & \rm \textbf{DS} \\
		\hline
		\(S_{1}\) & \rm A, C, C, A, T, G, T \\
		\hline
		\(S_{2}\) & \rm T, A, G, A, A, C, C \\
		\hline
		\(S_{3}\) & \rm A, G, G, C, A, A, T, C, T, C \\
		\hline
		\(S_{4}\) & \rm T, C, G, C, T, T, G, A, A, G \\
		\hline
		\(S_{5}\) & \rm T, A, A, G, G, A, C \\
		\hline
	\end{tabular}
\end{table}

\begin{strategy}[Sequences Pre-Read Pruning (SPRP) Strategy]
    \label{strategy:SPRP}
    \rm Before generating the patterns and building the Nettree, an operation is required, which is the reading of the sequence or sequence database. Before reading the sequences into the processing queue, we can filter some sequences which have no hope to generate the target frequent patterns, and thus we propose the strategy for pre-reading sequences named SPRP. When reading each sequence, if the query sequence \textit{qs} is not the sub-sequence of this reading sequence, then it is impossible to generate frequent target patterns from this sequence and hence there is no impact on the support of frequent target patterns. Furthermore, with the gap and length constraints, the support denoted as \textit{presup} can be pre-calculated by NETGAP algorithm with the \textit{gap} = [\textit{mingap}, \textit{maxlen} - \textit{qs.size}], where \textit{qs.size} represents the number of patterns in \textit{qs}. If \textit{presup} is equal to 0, the reading sequence is hopeless to generate the frequent target sequences.
\end{strategy}

For instance, \textit{qs} is set to $<$T, C$>$ in Table \ref{table3}. The gap constraints \textit{gap} in this example is set to [0, 4], and \textit{len} is set to [1, 4]. $S_1$ = "ACCATGT" does not contain the \textit{qs} = $<$T, C$>$, i.e., \textit{qs} is not a sub-sequence of $S_1$. $S_1$ is impossible to generate frequent target patterns and affects the calculation of support. Furthermore, the \textit{presup} is equal to 0 as well. Accordingly, the sequence $S_1$ is removed from the database $\mathcal{D}$. The \textit{presup} in $S_2$ is 0. Thus, the sequence $S_2$ is removed as well. It can be seen that this method of pre-calculating support is more suitable for gap-constrained sequential pattern mining than judging whether \textit{qs} is simply a sub-sequence of the reading sequence.

To explain the matching position in the sequence more clearly, the definitions of item match position of sequence and query item match position are, inspired by the previous work \cite{huang2022taspm}, as follows:

\begin{definition}[Query item matching position \rm \cite{huang2022taspm}]
    \label{definition 9}
    \rm The concept of the current query item, denoted as \textit{qi}, represents the matching item at the corresponding position of \textit{qs}. Its position is denoted as $q_{match}$. As the sequence match query items in \textit{qs}, a flag should be used to record the position of the corresponding item of the sequence. It is represented by item match position and denoted as $i_{match}$. The value of $i_{match}$ is actually the corresponding index of the sequence, and the value of $q_{match}$ is the index of $qs$. The matching position is recorded by the increasing of $i_{match}$ and $q_{match}$, which are both increased with the matching process. In this process, $i_{match}$ always increases, while $q_{match}$ only increases when it can match the corresponding item. If $q_{match}$ is equal to the size of \textit{qs}, it means that matching is completed. It is used in the strategy when sequence reading.

    Similarly, as the pattern matches query items in \textit{qs}, the value of $p_{match}$ is used to represent the corresponding index of the pattern, and the value of $q_{match}$ is the index of \textit{qs}, where the item with the index $q_{match}$ of the pattern is \textit{qi}, which means the first item that is not matched. The matching of \textit{qs} in the sequence is used to check whether the query pattern completely contains the query sequence, or which item of the query sequence the pattern corresponds to. The sequence from \textit{qi} to the last item of \textit{qs} is denoted as \textit{qis}. In the same way, since the first version of the TALENT algorithm needs to pre-extend the pattern in the front end, the current pattern and \textit{qs} need to be scanned in the way from the back to the front. The first item that is not matched with the pattern is denoted as \textit{qj}, and the sequence from the first item of \textit{qs} to \textit{qj} is denoted as \textit{qjs}.
\end{definition}

For example, given a sequence $S$ = "ATCCGT" and query sequence = $<$C, G$>$, before matching, $i_{match}$ and $q_{match}$ are both assigned to 0, which indicates that $i_{match}$ and $q_{match}$ are corresponding to the position of 0. They are `A' and `C', respectively. Each position movement judges the matched corresponding items. After matching the `C' in $<$C, G$>$, $i_{match}$ = 3 and $q_{match}$ = 1. After matching the `G' in $<$C, G$>$, $i_{match}$ = 5 and $q_{match}$ = 2, which means that the matching is completed.

In addition, if \textit{qs} = $<$C, T$>$, current pattern $P$ = $<$A, C, G$>$. It needs to judge the matching position of the query sequence corresponding to the pattern. if $q_{match}$ = \textit{qs.size}, it means that the matching is completed. However, if $q_{match}$ \textless $\textit{qs.size}$ but $q_{match}$ = \textit{pattern.size}, it means that matching is not complete and \textit{qi} is the first item that is not matched the pattern.

\begin{strategy}[Items Pre-Read Pruning (IPRP) Strategy]
    \label{strategy:IPRP}
    \rm In addition, a strategy named IPRP for pruning the front and back items in a sequence is proposed. In the process of matching the sequence with \textit{qs} from front to back, if $i_{match}$ + (\textit{qs.size} - $q_{match})$ $>$ \textit{maxlen}, the first $i_{match}$ items of the sequence are removed. Similarly, if $i_{match}$ + (\textit{qs.size}- 1 - $q_{match}$) $>$ \textit{maxlen}, the last $i_{match}$ items of the sequence are removed.
\end{strategy}

For example, in Table \ref{table3}, $S_1$ and $S_2$ have already been pruned from $\mathcal{D}$ according to the SPRP strategy. The value of \textit{gap} is set to [0, 4] and \textit{len} is [1, 4]. $S_3$ is pruned to "AATCTC" and $S_4$ is pruned to "TCGCTT". And $S_5$ becomes an empty sequence after the pruning of the prefixes and the suffixes.

\begin{algorithm}
	\small
	\caption{The TALENT$_{\rm BFS}$ algorithm}
	\label{alg:TALENT_BFS}
	\LinesNumbered
	\KwIn{The minimum support threshold \textit{minsup}, the gap constraints \textit{gap}, the length constraints \textit{len}, a database $\mathcal{D}$, and a query sequence \textit{qs}.}
	\KwOut{The frequent target set $S_F$}
	set $S_F$ and candidate set \textit{cand}[] as an empty set\;
	perform the Sequences Pre-Read Pruning and Items Pre-Read Pruning; // \textbf{Strategies 1 and 2} \\
	set \textit{N} $\leftarrow$ the number of sequences in $\mathcal{D}$\;
	generate all frequent 1-patterns whose support $\ge$ \textit{minsup} \;
	\ForEach{\rm $p$ \textbf{in} 1-patterns}{
		\textit{sup}($p$,$\mathcal{D}$) $\leftarrow$ $\sum_{i=1}^{N}$ NETGAP($p$,$S_i$)\;
		\If{\textit{sup}($p$,$\mathcal{D}$) $\ge$ \textit{minsup}}{
			\textit{cand}[\textit{p.size}].add($p$)\;
			\If{\textit{qs} $\in$ p}{
				$S_F$.add($p$);
			}
		}

		\ForEach{\rm 1-pattern $p$ \textbf{in} 1-patterns}{
			\If{\textit{qs} $\in p$}{
				continue\;
			}
			calculate the cases of \textit{sup}($P_E$) with the \textit{gap} = [\textit{mingap}, \textit{maxlen} - $P_E$.\textit{size}] and compare with \textit{minsup} in order\;
			\If{\rm all cases of \textit{sup}($P_E$) are all less than \textit{minsup}}{
				remove the $p$ from 1-patterns; \qquad // \textbf{Strategy 3}\\
			}
		}
	}
	$i$ $\leftarrow$ 1\;
	\While{\rm \textit{cand}[$i$] is not empty}{
		generate each pattern $p$ with size = $i + 1$ by calling \textit{gen\_candidate}\;		  \ForEach{\rm pattern $p$ \textbf{in} \textit{cand}[$i$]}{
			\If{\textit{qs} $\in p$}{
				continue\;
			}
			calculate the cases of \textit{sup}($P_E$) with the \textit{gap} = [\textit{mingap}, \textit{maxlen} - $P_E$.\textit{size}] and compare with \textit{minsup} in order\;
			\If{\rm all cases of \textit{sup}($P_E$) are all less than \textit{minsup}}{
				remove the $p$ from \textit{cand}[$i$]; \qquad // \textbf{Strategy 3}\\
			}
		}
		\textit{cand}[$i$ + 1] $\leftarrow$ \textit{gen\_candidate}(\textit{cand}[$i$])\;
		\ForEach{\rm $p$ \textbf{in} \textit{cand}[$i$ + 1]}{
			\If{\rm \textit{sup}($p$,$\mathcal{D}$) $\ge$ \textit{minsup} \&\& \textit{qs} $\in$ $p$ \&\& $S_F$ does not contains $p$}{
				$S_F$.add($p$);
			}
		}
		$i$ $\leftarrow$ $i$ + 1\;
	}
	\textbf{return} $S_F$
\end{algorithm}

After reading the sequences in the database, the generation of patterns is conducted. In the task of targeted non-overlapping sequential pattern mining, the pattern generation process of traditional algorithms toward non-overlapping sequential patterns will generate some redundant patterns. In order to meet this challenge, a pruning strategy used for breadth-first pre-extend named BPEP (Breadth-First Pre-Extend Pruning Strategy) is proposed.

\begin{figure}[ht]
    \centering
    \includegraphics[clip,scale=0.17]{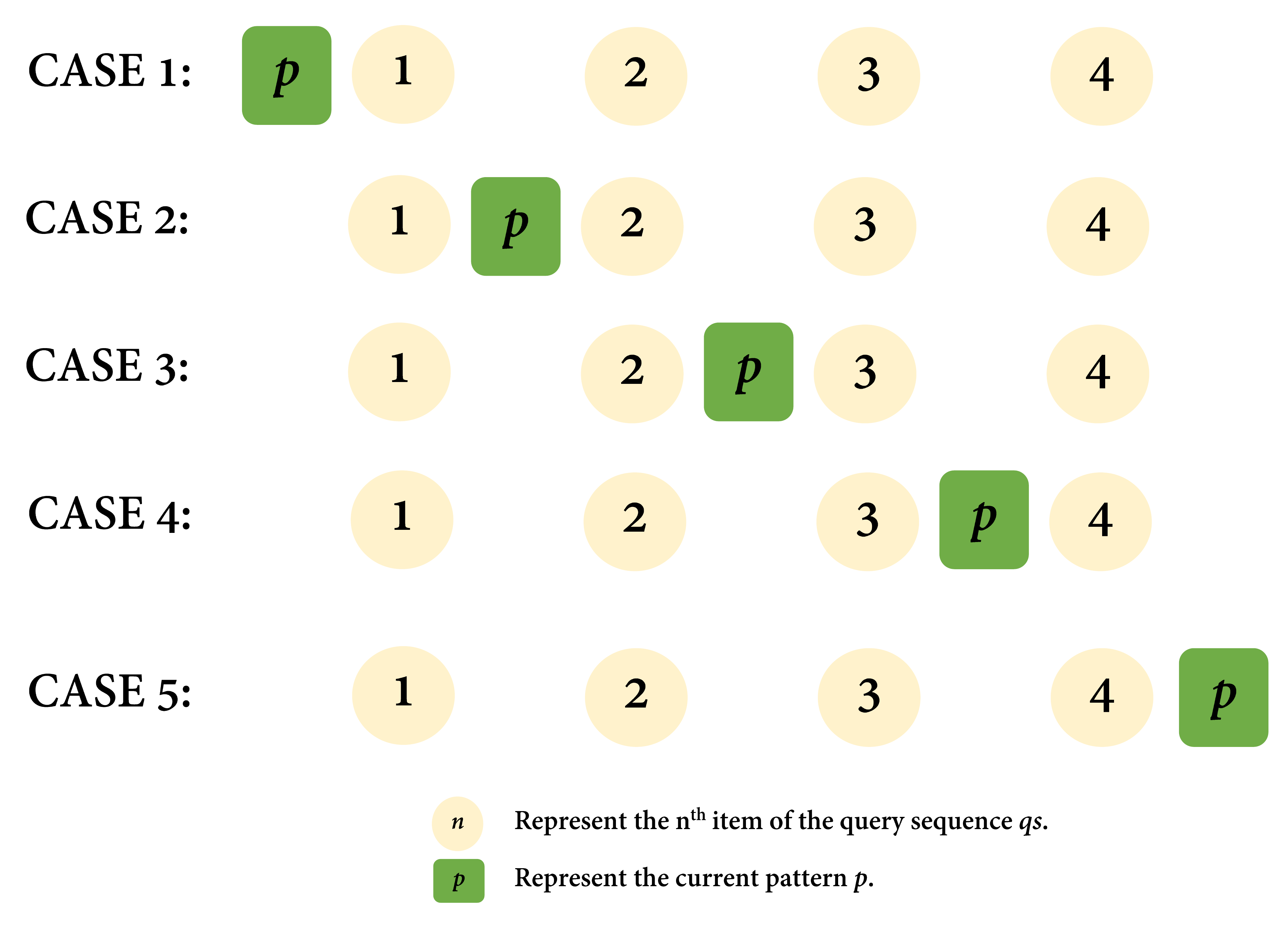}
    \caption{The cases that pattern \textit{p} is possible to be extended to a pattern which contains \textit{qs}.}
    \label{fig:strategyCase}
\end{figure}

\begin{figure}[b]
    \centering
    \includegraphics[clip,scale=0.22]{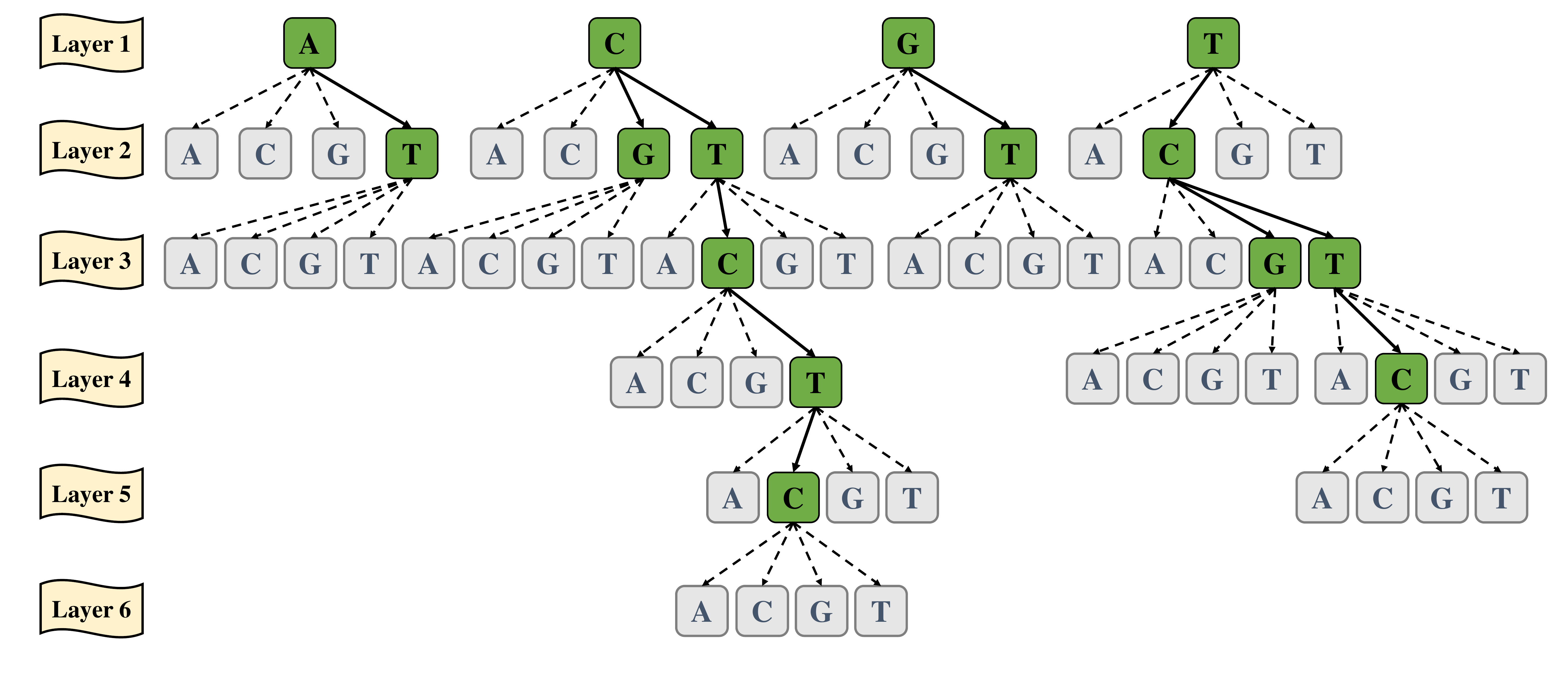}
    \caption{The process of pattern generation in DFS way.}
    \label{geneate_DFS}
\end{figure}

\begin{strategy}[Breadth-First Pre-Extend Pruning (BPEP) Strategy used in breadth-first search version]
    \label{Strategy3}
    \rm As mentioned above, the size of patterns is increasing one by one in the pattern generation process of breadth-first search. Hence, we propose a strategy to pre-extend the pattern to judge whether its super-patterns are expected to become a frequent target pattern, which is not used in the case that the current pattern contains query sequence \textit{qs} because the current pattern is already a frequent target pattern. We will discuss the cases when a pattern is possible to extend to a frequent pattern that contains \textit{qs}. As shown in Fig. \ref{fig:strategyCase}, assuming that the size of \textit{qs} is 4, the current pattern $p$ is possible in 5 positions to extend as a frequent pattern. For example, the current pattern $p$ is $<$A, C$>$, and the query sequence is $<$G, T$>$. Then the possible pre-extend cases are respectively $<$A, C, G, T$>$, $<$G, A, C, T$>$ and $<$G, T, A, C$>$. The number of possible cases is equal to \textit{qs.size} + 1. Actually, some situations need to be considered when the current pattern \textit{p} has already provided some item that becomes a part of \textit{qs}. Thus, the definitions of \textit{qi}, \textit{qj}, \textit{qis}, and \textit{qjs} in \ref{definition 8} are utilized in this strategy. To clearly state the matching process, we give an example like case 1 and case 5 in Fig. \ref{fig:strategyCase}. For instance, given the \textit{qs} = $<$A, T, C, C$>$ and \textit{p} = $<$G, A, T$>$, using the \textit{qj} to match unmatched item of \textit{qs} in $p$ form back to the front in case 1. The first item `C' is unmatched, and thus all items are spliced to the front end. $<$G, A, T$>$ is pre-extend to $<$A, T, C, G, A, T, C$>$. In case 5, the \textit{qi} matches the unmatched item of \textit{qs} in $p$ from front to back. The item `A' and `T' are matched, thus splice the $<$C, C$>$ to the back end of the \textit{p} to the pre-extended pattern $<$G, A, T, C, C$>$. Then, like in cases 2 to 4, judging needs to be made in two directions. The left part of \textit{qs} needs to be scanned from back to front like case 1 while the right part needs to be scanned from front to back like case 5. For case 2, $p$ needs to be placed between $<$A$>$ and $<$T, C, C$>$. Because in $p$ = $<$G, A, T$>$, as items `A' and `T' are respectively matched in two directions, the front end no longer needs to be spliced, and the back end needs to be spliced with $<$C, C$>$. Thus, the pre-extended pattern is $<$G, A, T, C, C$>$ in case 2. After discussing all the cases of pre-extension, we discuss how to determine whether to prune the current pattern. After every case of pre-extended pattern which is denoted as $P_E$ is generated, the support of $P_E$ is calculated by the NETGAP algorithm. It should be noted that the maximum gap constraints \textit{maxgap} should use the value of \textit{maxlen} - $P_{E}$.\textit{size} because the original \textit{gap} in the pre-extension situation may lead to omissions of some possible patterns. If the \textit{sup}($P_E$) is larger than \textit{minsup}, the current pattern $p$ is not to be pruned, and other cases are not necessary to be considered. Otherwise, it should be proceeded to calculate the support of $P_E$ in the next case. If $P_E$ in all cases are all less than \textit{minsup}, the current pattern should be removed from the candidate set.
\end{strategy}

With these three strategies and the post-processing technique, we proposed an effective version of the TALENT algorithm for non-overlapping sequential patterns. This version applies the idea of breadth-first search (BFS), which is shown as Algorithm \ref{alg:TALENT_BFS}.

\subsection{The Depth-First Search Version of TALENT Algorithm}
\label{TALENT_DFS}

The breadth-first search (BFS) version of the TALENT algorithm can achieve effective pruning operations, but pruning operations are a bit complicated. Hence, we develop another version of TALENT based on the idea of depth-first search (DFS), which enables an easy operation of pre-extension and pruning. As stated in the above context, in addition to the reading of the sequences, TALENT is divided into two main steps, which are respectively calculation of support and the generation of patterns. The support of patterns is calculated by the NETGAP algorithm, and the generation of patterns is stated in the following context.

\begin{algorithm}[h]
	\small
	\caption{The gen\_candidate\_DFS algorithm}
	\label{alg:candidate_DFS}
	\LinesNumbered
	\KwIn{The current pattern $p$, the minimum support threshold \textit{minsup}, the gap constraints \textit{gap}, the length constraints \textit{len}, a database $\mathcal{D}$, and a query sequence \textit{qs}.}
	\KwOut{generated patterns set $S_P$}

	\textit{sup}($p$,$\mathcal{D}$) $\leftarrow$ $\sum_{i=1}^{N}$ NETGAP($p$,$S_i$)\;
	\If{\textit{sup}($p$,$\mathcal{D}$) $\ge$ \textit{minsup}}{
		$S_F$.add($p$)\;
		get the matching position $q_{match}$\;
		\textbf{loop:}\\
		\If{\rm $p$ doesn't completely match \textit{qs}}{
			\textit{qis} $\leftarrow$ $<$\textit{qs}[$q_{match}$], \textit{qs}[$q_{match}$ + 1], $\cdots$, \textit{qs}[\textit{qs.size - 1}]$>$\;
			$P_E \leftarrow p \oplus$ \textit{qis}\;
			calculate the support of $P_E$ with \textit{newgap} = [\textit{mingap}, \textit{maxlen - qs.size}]\;
			\If{\textit{sup}($P_E$) \textless \textit{minsup}}{
				break \textbf{loop}; \qquad // \textbf{Strategy 4}
			}
		}
		save the pattern $p$\;
		\textit{pc} $\leftarrow$ the next item for extension\;
		$p$.UpdateTo($p \oplus$ \textit{pc})\;
		call gen\_candidate\_DFS($p$)\;
	}
	\textbf{return} $S_P$
\end{algorithm}

According to the domain of containing items, the pattern traversing layer by layer according to the size is no longer used in DFS version. Its detailed process is as follows. Firstly, start with 1-patterns in dictionary order. If a 1-pattern is frequent, continue to splice the items in $k$-pattern into ($k$ + 1) - patterns according to the dictionary order. If the pattern is infrequent, instead of continuing the depth traversal, splicing the item of next 1-patterns in the previous layer according to the dictionary order. We will continue to explain the process of pattern generation in DFS through the previous examples as shown in Fig. \ref{geneate_BFS}.

For example, given a sequence $S$ = "GTCAAGTCTCTCAGGT" in Table \ref{table2} with the condition of \textit{gap} = [0, 3], \textit{len} = [1, 8] and \textit{minsup} = 3 without query sequence \textit{qs}, this process is essentially implemented using the strategy of backtracking which is shown in Fig. \ref{geneate_DFS}. These steps are carried out after reading the data in the sequences of the database. Firstly, the 1-pattern $<$A$>$ is generated, and the support of which is calculated to judge whether it is a frequent pattern and whether it needs to be extended. Since its support is not less than \textit{minsup}, as the first super-pattern of $<$A$>$ according to dictionary order, pattern $<$A, A$>$ continues to do the same operation as above. If the support of the pattern exceeds or is equal to the \textit{minsup}, the operation of the extension will continue. Otherwise, return to the previous layer to extend the next pattern according to dictionary order. In this way, until all the items are mined. As shown in Fig. \ref{geneate_DFS}, all frequent patterns mined in order are respectively $<$A$>$, $<$A, T$>$, $<$C$>$, $<$C, G$>$, $<$C, T$>$, $<$C, T, C$>$, $<$C, T, C, T$>$, $<$C, T, C, T, C$>$, $<$G$>$, $<$G, T$>$, $<$T$>$, $<$T, C$>$, $<$T, C, G$>$, $<$T, C, T$>$, and $<$T, C, T, C$>$. The above process is realized by recursively calling the function gen\_candidate\_DFS as shown in Algorithm \ref{alg:candidate_DFS}. It can be easily seen that the DFS version of non-overlapping sequential pattern mining has more calculation count, but we expect it to be better adapted to the proposed strategy so that it does not need too many operations to achieve more efficient targeted mining of non-overlapping patterns.

\begin{algorithm}[h]
	\small
	\caption{The TALENT$_{\rm DFS}$ algorithm}
	\label{alg:TALENT_DFS}
	\LinesNumbered
	\KwIn{The minimum support threshold \textit{minsup}, the gap constraints \textit{gap}, the length constraints \textit{len}, a database $\mathcal{D}$, and a query sequence \textit{qs}.}
	\KwOut{The frequent target set $S_F$}

	set $S_F$ as an empty set\;
	perform the sequence pre-read pruning; \qquad // \textbf{Strategy 1}\\
	perform the item pre-read pruning in each sequence; \qquad //  \textbf{Strategy 2} \\
	generate all frequent 1-patterns whose support $\ge$ \textit{minsup} \;
	\ForEach{\rm \textit{p} in frequent 1-patterns}{
		\If{\rm $p$ doesn't completely match \textit{qs}}{
			get the matching position $q_{match}$\;
			\textit{qis} $\leftarrow$ $<$\textit{qs}[$q_{match}$], \textit{qs}[$q_{match}$ + 1], $\cdots$, \textit{qs}[\textit{qs.size - 1}]$>$\;
			$P_E \leftarrow p \oplus$ \textit{qis}\;
			calculate the support of $P_E$ with \textit{newgap} = [\textit{mingap}, \textit{maxlen - qs.size}]\;
			\If{\rm \textit{sup}($P_E$) \textless \textit{minsup}}{
				break \textbf{foreach}; \qquad // \textbf{Strategy 4}
			}
		}
		\textit{sup}($p$,$\mathcal{D}$) $\leftarrow$ $\sum_{i=1}^{N}$ NETGAP($p$,$S_i$)\;
		\If{\rm \textit{sup}($p$,$\mathcal{D}$) $\ge$ \textit{minsup} \&\& \textit{qs}$\in p$}{
			$S_P$ = call \textit{gen\_candidate\_DFS}($p$)\;
			$S_F$.append($S_P$)\;
		}
	}
	\textbf{return} $S_F$
\end{algorithm}

After the query sequence \textit{qs} is added into the task to mine the frequent target patterns, similar to the BFS version, a post-processing technique is applied to select out the frequent target patterns which contain \textit{qs} as well. In addition, the strategy SPRP and IPRP are also valid in the DFS version of TALENT. However, for the reason that the way of pattern generation is different from the other version, the Depth-First Pre-Extend Pruning Strategy briefly named DPEP is used to reduce the number of redundant patterns before pattern extension. It is stated as follows.

\begin{strategy}[Depth-First Pre-Extend Pruning (DPEP) Strategy used in depth-first search version] \rm Before the extension of a pattern whose support is not less than \textit{minsup}, this strategy can be used to determine whether the extended patterns of it is possible to become frequent target sequences which contain the query sequence \textit{qs}. The pre-extension of a pattern is supplementing items at the end of a pattern to satisfy the condition of containing the query sequence. Let \textit{qis} = $<$\textit{qs}[$q_{match}$], \textit{qs}[$q_{match}$ + 1], $\cdots$, \textit{qs}[\textit{qs.size - 1}]$>$, the above step is actually supplementing the current pattern with \textit{qis}. The pre-extended pattern is denoted as $P_{E}$. The support of $P_E$ are calculated with \textit{gap} = [\textit{mingap}, \textit{maxlen} - \textit{qs.size}] and \textit{len} = [\textit{minlen}, \textit{maxgap}]. If the \textit{sup}($P_E$) is less than \textit{minsup}, all super-patterns of $P$ are impossible to be frequent target patterns. In other words, pattern $P$ is no need to continue the search in depth.
\end{strategy}

For instance, for the sequence $S$ = "GTCAAGTCTCTCAGGT" in Table \ref{table2} with the condition of \textit{gap} = [0, 3], \textit{len} = [1, 8], \textit{minsup} = 3 and query sequence \textit{qs} = $<$T, C$>$, although the support of pattern $<$G$>$ is larger than \textit{minsup}, the support of pre-extended pattern $P_E$ = $<$G, T, C$>$ is satisfying target but not frequent. The pattern $<$G$>$ does not continue the extension because all of its super-patterns are impossible to be frequent target patterns.

According to what is stated above, the complete pseudocode of the DFS version of TALENT is shown in Algorithm \ref{alg:TALENT_DFS}. The strategy DPEP is used in the generation of patterns in the DFS version, both in the recursion in Algorithm \ref{alg:candidate_DFS} and the generation of 1-patterns in Algorithm \ref{alg:TALENT_DFS} to end the extension of hopeless patterns so that the calculation number of redundant patterns are reduced.

\section{Experiments}  \label{sec:experiments}

In order to assess the performance and memory usage of the two versions of the TALENT algorithm, a series of experiments were performed as follows: TALENT was compared to NOSEP$_{\rm Ta}$, which uses post-processing to filter discovered patterns containing query sequences. Except for \textit{minsup}, there are still some parameters that can affect the performance of the proposed algorithm, such as the length of the query sequence. Thus, we also evaluate the effect of the length of \textit{qs} in TALENT.

All the experiments are carried out on a 64-bit Windows 11 personal computer equipped with a 12$\rm ^{th}$ Gen Intel (R) Core™ i7-12700F CPU and 16 GB of RAM. Algorithms are implemented in Java using Eclipse IDE 2022. For the reproducibility requirement, the source code and datasets are available at GitHub\footnote{https://github.com/DSI-Lab1/TALENT}. The details of the experimental datasets are as follows.

\subsection{Datasets for the Experiments}

We use nine datasets in total, which are used to carry out the following experiments. Among them, six biological information datasets are collected from the National Center for Biotechenique Information (NCBI)\footnote{https://www.ncbi.nlm.nih.gov/labs/virus/vssi/}, and other three datasets are acquired from SPMF\footnote{https://www.philippe-fournier-viger.com/spmf/index.php?link=datasets.php}. The sequences from a kind of virus are collected and made into a dataset. These datasets are formatted as sequences suitable for the algorithm to mine the sequential patterns with gap constraints. To facilitate classification and discrimination, they are named \textit{MERS\_DNA}, \textit{Ebola\_DNA}, \textit{Zaki\_DNA}, \textit{Dengue\_Protein}, \textit{FCoV\_Protein} and \textit{Rota\_Protein}. As the name suggests, among these six biological datasets, three are DNA sequence datasets, and the other three are protein datasets. Each item represents the nucleotide species in DNA and the amino acid species in proteins.

\begin{itemize}
    \item  \textbf{MERS\_DNA}. This dataset is collected from the DNA sequence, fragments of humans, or other primates infected with the Middle East Respiratory Syndrome (MERS) caused by the Middle East Respiratory Syndrome Coronavirus (MERS-CoV).
 
    \item  \textbf{EBHF\_DNA}. This dataset is collected from the DNA sequence, fragments of humans, or other primates infected with Ebola hemorrhagic fever (EBHF) caused by the Ebola virus.
 
    \item  \textbf{Zaki\_DNA}. This dataset is collected from the DNA sequence, fragments of humans, or other primates infected with the Zaki virus disease caused by the Zika virus (ZIKV).
 
    \item  \textbf{Dengue\_Protein}. This dataset is collected from the sequence of proteins, fragments of peptide chains of humans, or other primates infected with Dengue Fever (DF) caused by the dengue virus or vector insects such as Aedes ae-gyPti and aedes-albopictus.
 
    \item  \textbf{FCoV\_Protein}. This dataset is collected from the sequence of proteins, and fragments of peptide chains of cats, or other felines infected with intestinal diseases caused by the Feline Coronavirus (FCoV).
 
    \item  \textbf{Rota\_Protein}. This dataset was compiled from the protein sequences and peptide chain fragments of human infants infected with rotavirus or virus-host diarrhea.
\end{itemize}

Actually, the TALENT algorithm is suitable for processing non-biological datasets because there is only one letter in each item of these datasets. In order to evaluate algorithm performance more comprehensively and broaden the possible application range of the algorithm, three non-biological datasets are used in the experiments, which are respectively named BIKE, Leviathan, and BMSWebView2.

\begin{itemize}
    \item  \textbf{BIKE}. This dataset contains location sequences of shared bikes parked in one city. Each item represents a shared bike's docking station, and each sequence represents a different location point where a shared bike stops over time. It was obtained originally from Andrea Tonon and transformed from Kaggle data\footnote{https://www.kaggle.com/cityofLA/los-angeles-metro-bike-share-trip-data}. 
	
    \item  \textbf{Leviathan}. This dataset is a conversion of the novel \textit{Leviathan}\footnote{https://en.wikipedia.org/wiki/Leviathan} by Thomas Hobbes which was published in 1651. It is a sequence dataset, where each item means one word.
	
    \item  \textbf{BMSWebView2}. This is a dataset from the KDD CUP 2000. It contains click-stream data from an e-commerce website, and each item of that data means a click event.
\end{itemize}

Table \ref{table:Datasets feature} shows the characteristics of the datasets used in the experiments. Note that \textit{TL} is the total length of a dataset, \textit{NoI} is the number of items a dataset contains, \textit{NoS} is the number of sequences a dataset contains, and \textit{ALoS} is the average length of each sequence in the dataset.

\begin{table}[h]
    \caption{The features of datasets in experiments}
	\label{table:Datasets feature}
	\centering
	\begin{tabular}{|c|c|c|c|c|}
		\hline
		\rm \textbf{Sequence Dataset} & \rm \textbf{\textit{TL}} & \rm \textbf{\textit{NoI}} & \rm \textbf{\textit{NoS}} & \rm \textbf{\textit{ALoS}}  \\ \hline   \hline     
		\rm \textbf{MERS\_DNA} & 49,288 & 4 & 165 & 298.7  \\ \hline
		\rm \textbf{EBHF\_DNA} & 19,220 & 4 & 259 &  74.2 \\ \hline
		\rm \textbf{Zaki\_DNA} & 56,834 & 4 & 288 &  197.3 \\ \hline
		\rm \textbf{Dengue\_Protein} & 524,677 & 21 & 5,302 & 99.0 \\ \hline
		\rm \textbf{FCoV\_Protein} & 190,221 & 21 & 2,585 & 73.6 \\ \hline
		\rm \textbf{Rota\_Protein} & 326,918 & 21 & 3,848 & 85.0  \\ \hline
		\rm \textbf{BIKE} & 153,383 & 67 & 21,078  & 7.3  \\ \hline
		\rm \textbf{Leviathan} & 197,251 & 5,834 & 9,025 & 33.8  \\ \hline
		\rm \textbf{BMSWebView2} & 358,278 & 3,340 & 77,512 & 4.6  \\ \hline	
	\end{tabular}
\end{table}

\subsection{Efficiency Analysis} \label{subsection:efficiency}

The parameters that affect the efficiency of the mining process include gap constraint \textit{gap}, length constraint \textit{len}, minimum support threshold \textit{minsup}, and query sequence \textit{qs}. To ensure the consistency of other variables in the experiment, the gap and length constraints are set to \textit{gap} = [0, 3] and \textit{len} = [1, 10] in the following experiments. In addition, \textit{minsup} and \textit{qs} will be set to different values and sequences in different datasets.

\begin{figure}[h]
    \centering
    \includegraphics[clip,scale=0.45]{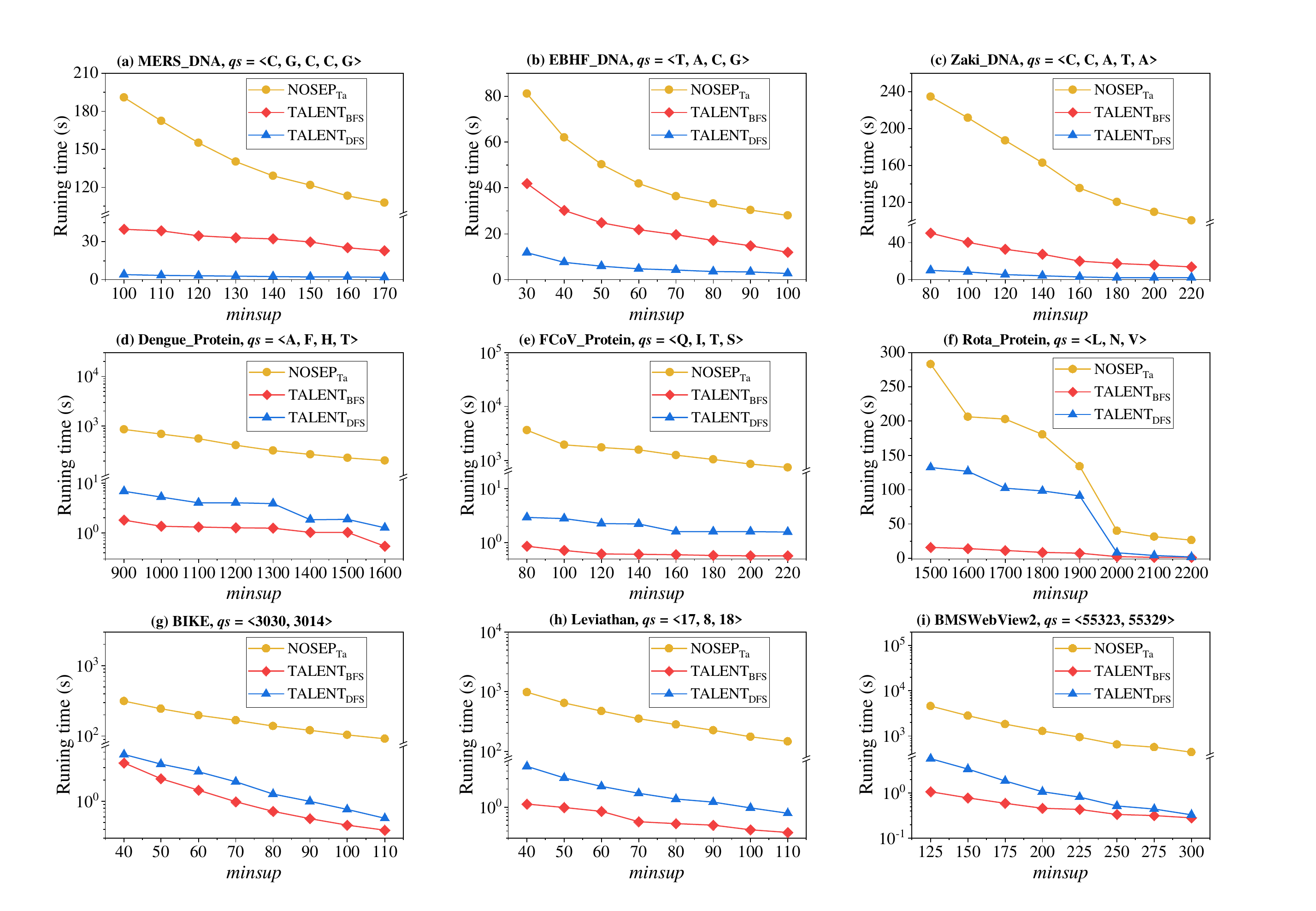}
    \caption{Running time of various datasets in different values of \textit{minsup}.}
    \label{fig:RunningTime}
\end{figure}

In the following, we carry out several experiments to show the efficiency of various datasets in the three algorithms, and the running time is used to quantify the efficiency. The baseline is NOSEP$_{\rm Ta}$, which is extended by the NOSEP algorithm \cite{wu2017nosep}. On the basis of NOSEP, post-processing techniques are added to filter out the target patterns. TALENT$_{\rm BFS}$ and TALENT$_{\rm DFS}$ are two versions of TALENT. After the comparison of experiments, the running time of three algorithms is shown in Fig. \ref{fig:RunningTime}.

\begin{itemize}
    \item  With the increase of \textit{minsup}, all the running times of the three algorithms tend to decrease. It shows that both versions of the TALENT algorithm with different \textit{minsup} have different degrees of optimization compared to the baseline algorithm NOSEP$_{\rm Ta}$.
	
    \item The DNA datasets are used in Fig. \ref{fig:RunningTime} (a) to (c). It shows that the running time of the BFS version of TALENT is less than half that of the baseline. Further, the DFS version of TALENT shows better efficiency. The running time is one to two orders of magnitude less than the baseline. The above experiments indicate that the DFS version of TALENT shows superior efficiency in the task of targeted non-overlapping sequential pattern mining in the DNA dataset.
	
    \item In Fig. \ref{fig:RunningTime} (d) to (f), the protein datasets are utilized to conduct the experiments. The results of the experiments show that both versions of TALENT perform better than the baseline algorithm. Contrary to the above result about the optimization effect of the two versions of the algorithm, the BFS version of TALENT shows superior efficiency to the DFS version in the task of targeted non-overlapping sequential pattern mining in DNA datasets. In Dengue\_Protein, the runtime of the DFS version of TALENT is around two orders of magnitude less than the baseline, and the runtime of the BFS version is about three orders of magnitude less than the baseline. In the FCoV\_Protein dataset, both versions of the algorithm are an order of magnitude faster than in the previous dataset compared with the baseline algorithm. In Rota\_Protein, the runtime of the DFS version of TALENT is less than half that of the baseline, and the runtime of the BFS version is one to two orders of magnitude less than the baseline.
	
    \item In Fig. \ref{fig:RunningTime} (g) to (i), the high efficiency of TALENT is also reflected in three non-biological datasets. It shows that TALENT has different optimization performances in different datasets, but the BFS version performs better than the DFS version in all non-biological datasets because the number of items is usually larger than that of biological datasets. It can be proved that TALENT is suitable for non-biological datasets as well as experimentally, which shows that TALENT may have better application value in more application scenarios.
\end{itemize}

We can draw the conclusion that, under the same conditions, the DFS version shows higher efficiency in the DNA dataset, while the BFS version shows higher efficiency in the protein dataset. The reason for this situation is mainly the number of items. Due to the fact that the number of items in the DNA sequence is only 4, through the DFS version of the algorithm, a frequent target pattern with the support of more than \textit{minsup} and containing a query sequence \textit{qs} is only needed to traverse four times each time. The traversal process is quite simple with the pre-extension operation of DFS. Compared with the complex pre-extension process and connection operations of prefixes and suffixes in BFS, the running time of the DFS version of TALENT is, in all likelihood, shorter. However, the situation changes in the protein dataset. Since there are 21 kinds of amino acids in a protein, the number of items in the protein datasets is 21. Thus, to conduct the extension, every frequent pattern needs to traverse the number of frequent items in the 1-pattern set, which is up to 21. The number is relatively large, thus the connection operations of prefixes and suffixes in BFS play a significant role. Although the operations of pre-extension and connection are complete, it can filter much more patterns than when the number of items is only four. Therefore, the efficiency of the DFS version is higher than the BFS version in the DNA dataset, and the efficiency of the BFS version is higher than the DFS version in the protein dataset. And in the non-biological datasets, the number of items is usually larger than in DNA sequences and protein sequences, so the efficiency of the BFS version shows more effectiveness in these datasets.

\subsection{Memory Analysis}  \label{subsection:memory}

\begin{figure}[h]
    \centering
    \includegraphics[clip,scale=0.45]{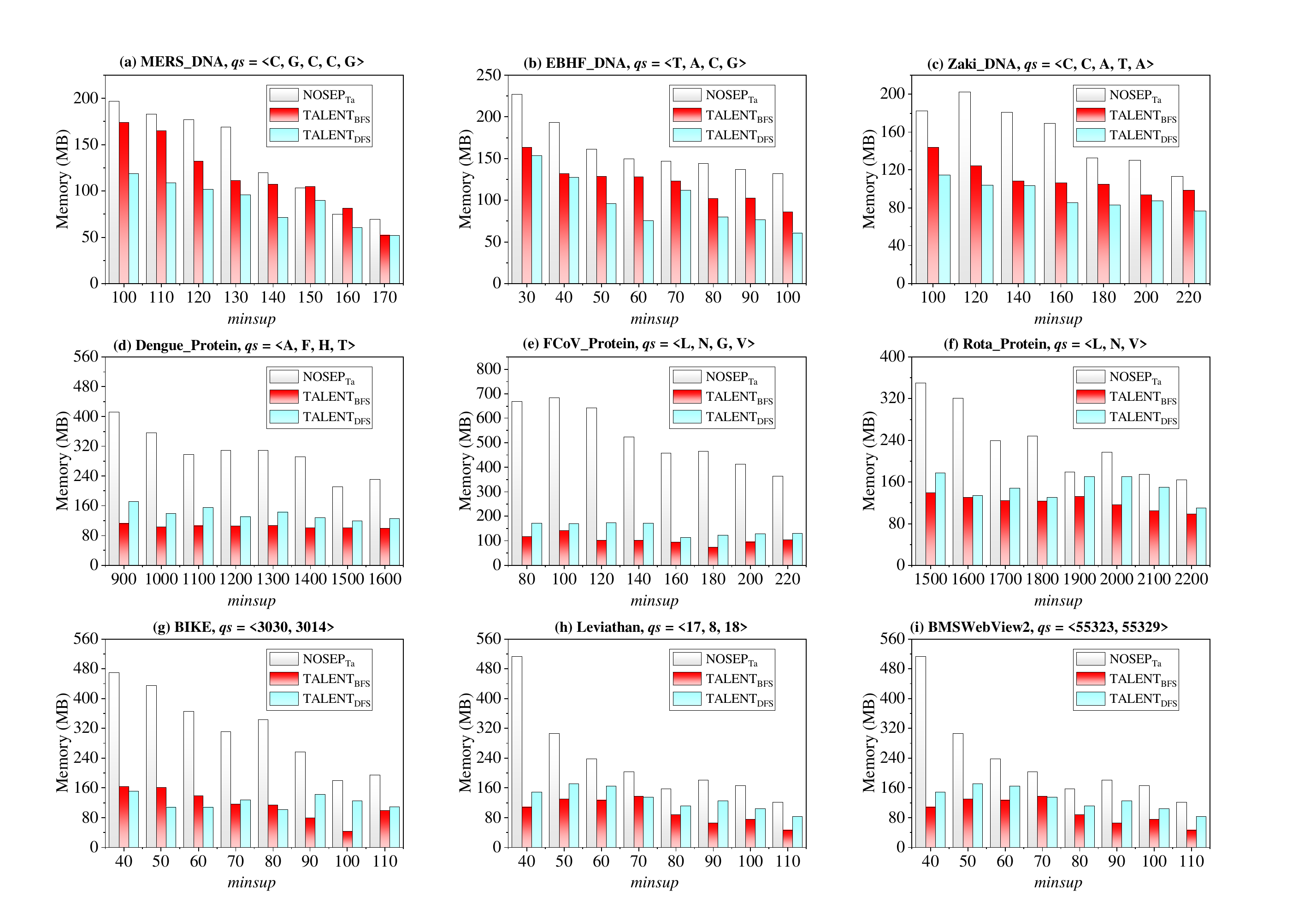}
    \caption{Using maximum memory of various datasets in different values of \textit{minsup}.}
    \label{fig:Memory}
\end{figure}

The consumption of memory is an important indicator of the algorithm's performance. Memory consumption depends on a lot of factors, including the total length of sequence datasets, the number of items, the average length of each sequence, etc. The memory consumption of two versions of TALENT and the baseline NOSEP$\rm _{Ta}$ are shown in Fig. \ref{fig:Memory}. With the increase of \textit{minsup}, memory consumption tends to decrease. Except for a few cases, we find that both versions of TALENT consume less memory than the baseline. It shows similar characteristics to running time in that the memory of the DFS version of TALENT consumes less than the BFS version in the DNA sequence, but the conclusion is the opposite in the protein sequence. Actually, the reason is the same as what we discussed in the previous section about why the efficiency of the two versions of TALENT varies across datasets. Because the BFS version of TALENT has a more dramatic effect on reducing the number of traversals, the pattern generation with the larger number of items obviously performs well in the BFS version. On the contrary, when the number of items is smaller, the operation of the BFS version is complex but can't filter a large number of items to be extended. Thus, in terms of two versions of the optimization effect, the memory consumption performs similarly to the runtime.

Furthermore, the memory consumption in NOSEP$\rm _{Ta}$ is heavily influenced by the datasets' characteristics. However, it diminishes under TALENT. Actually, this performance is inevitable. More items mean that more memory will inevitably be consumed, but it can be massively reduced with the strategies SPRP and IPRP. Hence, there is not much difference in memory consumption with different numbers of items. For example, in Fig. \ref{fig:Memory} (b) and (f), with the same length of \textit{qs}, the memory in EBHF\_DNA consumes approximately 150 MB to 230 MB, whereas the memory in FCoV\_Protein consumes approximately 350 MB to 700 MB in NOSEP$\rm _{Ta}$. And both versions of TALENT consume 50 MB to 175 MB in the two datasets. This aspect shows that TALENT performs better in datasets with multiple items. This is because, prior to reading the datasets, the SPRP and IPRP strategies are more likely to be effective on datasets with a larger number of items. When the number of items is small, it is difficult for the support of \textit{qs} to be zero because item occurrences are frequently very intense. Thus, the pruning number of sequences is relatively less. By the same token, the IPRP strategy makes it tougher to prune the items as well. In terms of memory consumption, the datasets with more items are more likely to have a greater improvement.

Based on the running time and memory consumption, the TALENT algorithm outperforms the baseline NOSEP$\rm _{Ta}$ algorithm. Moreover, for different datasets, the two versions of TALENT have different optimization effects.

\begin{figure}[b]
    \centering
    \includegraphics[clip,scale=0.3]{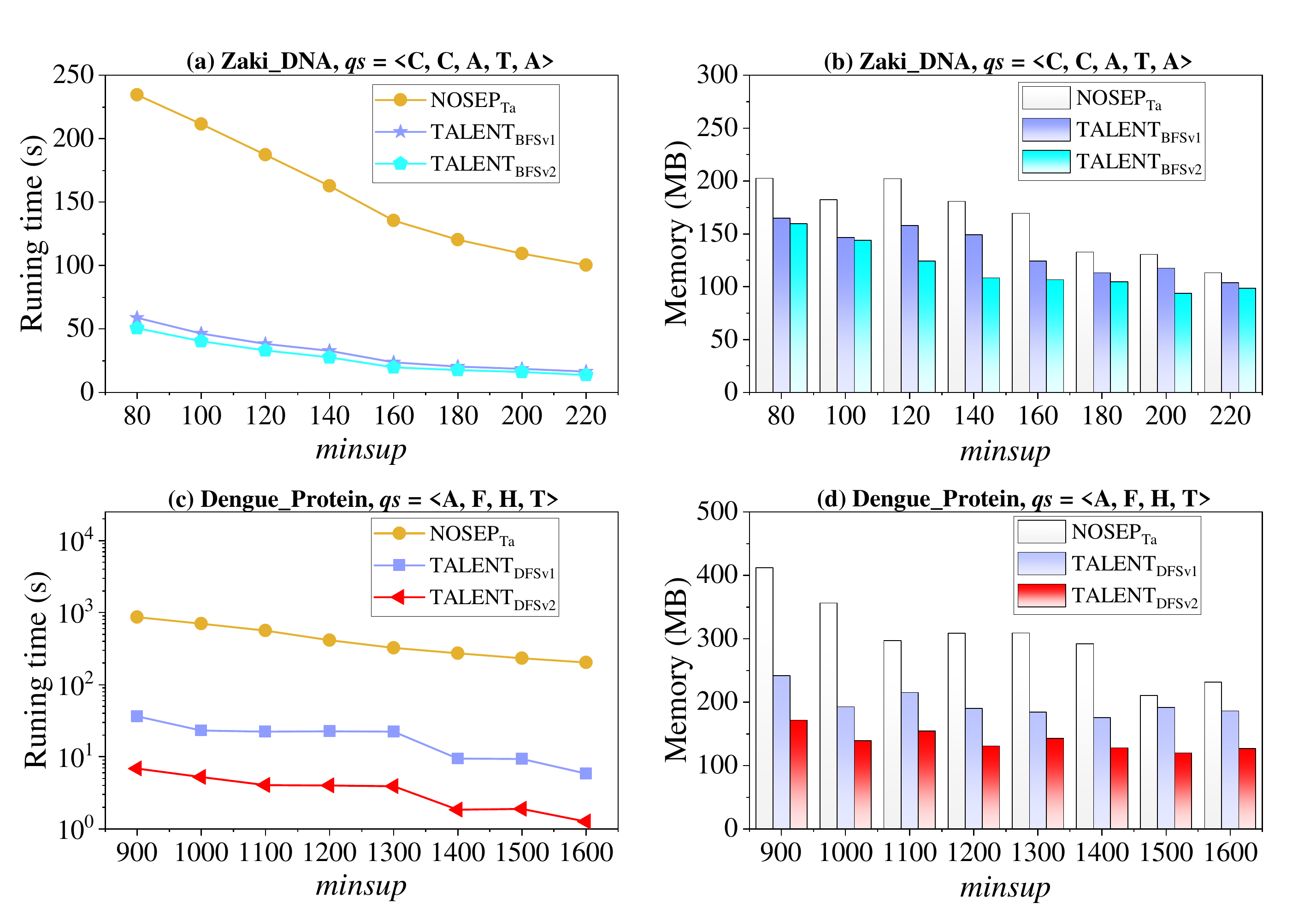}
    \caption{Running performance of the variants of two versions of the algorithm.}
    \label{fig:Strategy}
\end{figure}

\subsection{Strategy Evaluation}

These pruning strategies are mainly divided into two stages: the pre-reading pruning stage and the pre-extension pruning stage. The strategies SPRP and IPRP are in the pre-reading pruning stage, and the strategies BPEP and DPEP are in the pre-extension stage. Accordingly, we divide the BFS version of TALENT into two variants: TALENT$_{\rm BFS v1}$ (it only uses the BPEP strategy and the post-processing technique) and TALENT$_{\rm BFS v2}$ (it uses the post-processing technique and all strategies for TALENT$_{\rm BFS}$). Similarly, the DFS version is also divided into two variants: TALENT$_{\rm DFS v1}$ (it only uses the DPEP strategy and the post-processing technique) and TALENT$_{\rm DFS v2}$ (it uses the post-processing technique and all strategies for TALENT$_{\rm DFS}$). To conduct this experiment, we select the Zaki\_DNA dataset for the BFS version of TALENT and select Dengue\_Protein for the DFS version. In terms of both running time and memory consumption, TALENT$_{\rm BFS v1}$ has better performance than NOSEP$\rm _{Ta}$ in Fig. \ref{fig:Strategy} (a) and (b). Thus, the strategies of the BFS version of TALENT are all effective. Similarly, it can be proved in the experiment that the strategies of the DFS version of TALENT are all effective (see Fig. \ref{fig:Strategy} (c) and (d)). Therefore, from Fig. \ref{fig:Strategy}, it can be seen that the strategies for both pre-reading pruning and pre-extension pruning play a role in our effective algorithm TALENT. Especially, by comparison of TALENT$\rm _{BFS v1}$ and NOSEP$\rm _{Ta}$ as well as TALENT$\rm _{DFS v1}$ and NOSEP$\rm _{Ta}$, from Fig. \ref{fig:Strategy}, it can be seen that the efficiency improvement of the BPEP and DPEP strategies is very remarkable in different datasets.

Section \ref{subsection:memory} has demonstrated that datasets with more items are more likely to be improved due to the SPRP and IPRP strategies. It has been amply demonstrated experimentally. From Fig. \ref{fig:Strategy}, it can be seen that the number of items is relatively large, and the pre-reading strategy makes it easier to filter a series of sequences or items in the protein sequence dataset. For example, in terms of running time, in the Dengue\_Protein dataset shown in Fig. \ref{fig:Strategy} (b), the running time of TALENT with the pre-extension pruning strategy has improved by an order of magnitude, and the running time of TALENT with the pre-reading pruning strategy has improved by an order of magnitude as well. However, in the Zaki\_DNA dataset shown in Fig. \ref{fig:Strategy} (a), the running time of TALENT with the pre-extension pruning strategy has improved by an order of magnitude while the pre-reading strategy has improved by far less than an order of magnitude. That means that the datasets with more items are more likely to benefit from the pre-reading strategy, as described in the previous section.

In summary, all these proposed strategies are effective. The most critical pruning strategies are BPEP and DPEP, which are the key to TALENT$\rm _{BFS}$ and TALENT$\rm _{DFS}$, respectively. The SPRP and IPRP strategies are more efficient in a sequence database that contains more items.

\subsection{The Effect of Query Sequence Length}

\begin{figure}[h]
    \centering
    \includegraphics[clip,scale=0.4]{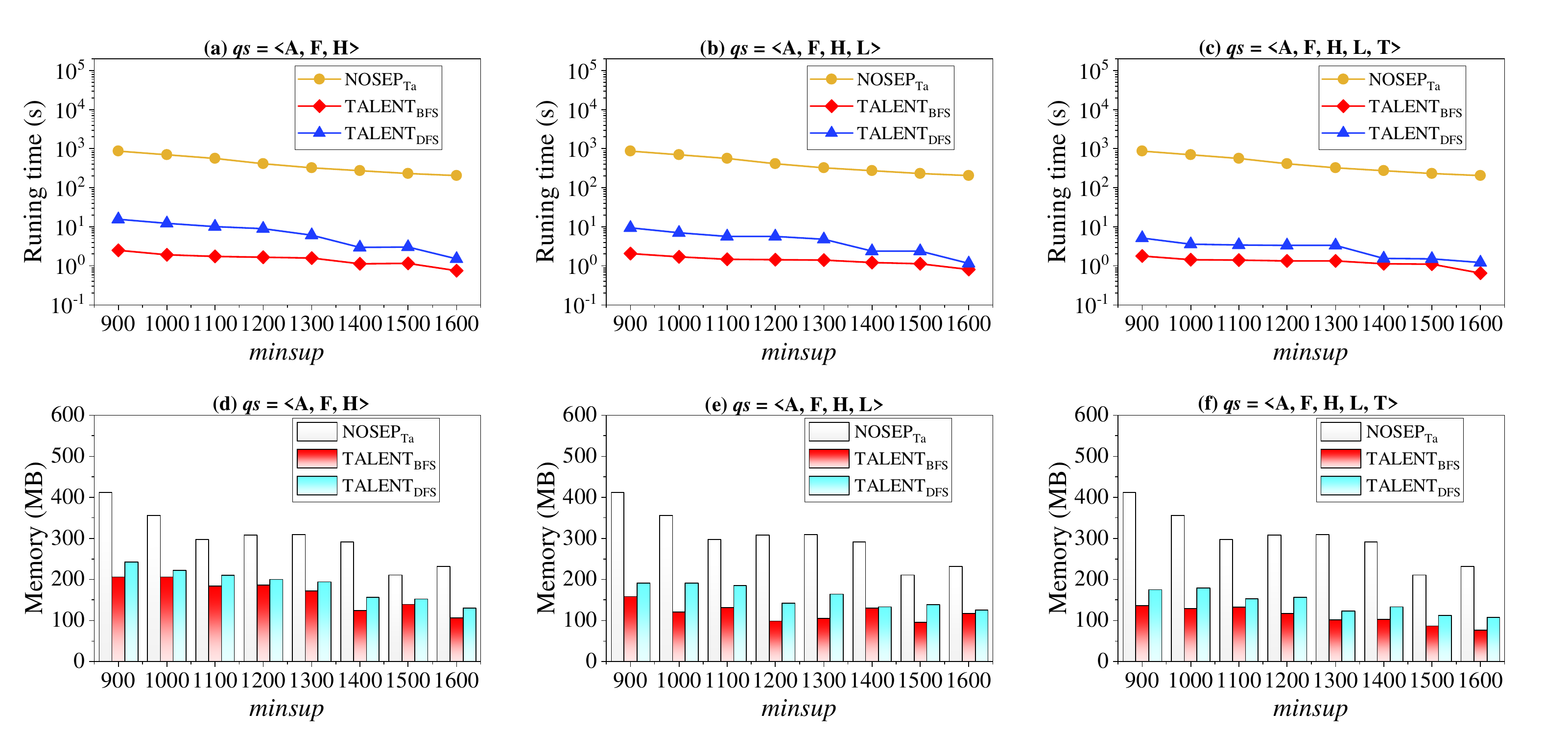}
    \caption{Running time and memory consumption in Dengue\_Protein.}
    \label{fig:Query}
\end{figure}

In this section, we discuss the effect of the length of the query sequence \textit{qs} on the TALENT algorithm. In order to exclude the influence of other irrelevant variables, we only use one dataset, Dengue\_protein, to conduct the experiment with the \textit{gap} = [0, 3] and \textit{len} = [1, 10]. In order to explore the effect of \textit{qs} length, we set the \textit{qs} as $<$A, F, H$>$, $<$A, F, H, L$>$ and $<$A, F, H, L, T$>$, respectively. Accordingly, the results of the experiment are shown in Fig. \ref{fig:Query}. The running time and memory consumption with various values of \textit{minsup} and different \textit{qs}.

\begin{table}[b]
	\centering
	\caption{Number of frequent target patterns in dataset Dengue\_Protein with various \textit{minsup} and \textit{qs}}
	\label{table5}
	\begin{tabular}{|c|c|c|c|c|c|c|c|c|}  
		\hline 
		\rm \textbf{\textit{minsup}} & \rm \textbf{900} & \rm \textbf{1000} & \rm \textbf{1100} & \rm \textbf{1200} & \rm \textbf{1300} & \rm \textbf{1400} & \rm \textbf{1500} & \rm \textbf{1600} \\
		\hline  
		\rm \textit{qs} = $<$A, F, H$>$ & \rm 116 & \rm 79 & \rm 75 & \rm 71 & \rm 61 & \rm 30 &\rm 30 & \rm 12 \\ 
		\hline
		\rm \textit{qs} = $<$A, F, H, L$>$ & \rm 53 & \rm 34 & \rm 33 & \rm 32 & \rm 30 & \rm 15 &\rm 15 & \rm 6 \\ 
		\hline
		\rm \textit{qs} = $<$A, F, H, L, T$>$ & \rm 29 & \rm 15 & \rm 15 & \rm 14 & \rm 14 & \rm 9 &\rm 8 & \rm 2 \\  
		\hline  
	\end{tabular}
\end{table}

As shown in Table \ref{table5}, the number of frequent target patterns decreases with the increase of the length of the query pattern, and the number of frequent target sequences decrease not only as the \textit{minsup} increases but also as the length of \textit{qs} increases. First, based on the results of the experiment, TALENT outperforms the baseline algorithm in all experiments with various \textit{minsup} and \textit{qs}. Moreover, Fig. \ref{fig:Query} (a) to (c) shows that the NOSEP$\rm _{Ta}$ does not change much as the length of a query pattern increases in terms of running time. Similarly, the same goes for memory consumption performance as shown in Fig. \ref{fig:Query} (d) to (f). However, the running time and memory consumption of TALENT is less when the length of \textit{qs} is shorter. After analyzing, TALENT can filter more patterns that are hopeless to become frequent target patterns when the length of \textit{qs} is larger than the case that the length of \textit{qs} is shorter. The experiments conducted to show this analysis is correct. Fig. \ref{fig:Query} shows that the running time and memory consumption both perform better when the \textit{qs} length is larger. In other words, the experiments show the characteristics of the algorithm, the longer the \textit{qs} length, the better its performance. The performance of different \textit{qs} further illustrates the value of its application. If only a few patterns are found, the algorithm will quickly terminate rather than continue to take more resources to search for meaningless and hopeless patterns like the baseline algorithm. In summary, TALENT$\rm _{DFS}$ is the better algorithm for the protein sequence dataset than TALENT$\rm _{BFS}$, and the performance of the algorithm will be optimized with the increase of \textit{qs} length and the decrease of valid patterns.

\section{Conclusions and Future Works}  \label{sec:conclusion}

In this paper, to address the problems of targeted sequential pattern mining with flexible gap constraints, the concept of target non-overlapping sequential pattern is defined. An efficient algorithm called TALENT is proposed to quickly discover more focused results. It makes more sense than traditional non-overlapping SPM for the reason that targeted mining is more specific, targeted, and valuable. In the TALENT algorithm, a post-processing technique and a total of four strategies are used to select the patterns that contain one or more query sequences and reduce meaningless calculations. Finally, experimental results show that TALENT outperforms the baseline NOSEP$\rm _{Ta}$ algorithm on all real datasets tested with different query sequences. Moreover, two versions of TALENT demonstrate their different strengths on different types of datasets.

In the future, we will address other meaningful mining tasks by extending TALENT, such as TALENT with top-$k$ patterns, TALENT via closed patterns, etc. These interesting tasks can reduce redundancy in the TALENT algorithm. Furthermore, based on targeted pattern mining, we would like to propose effective algorithms for specific application scenarios. For example, it is clear that targeted pattern mining from different types of data (e.g., sequence data, uncertain data, event-based data, and multi-source heterogeneous data) with more flexible gap constraints under different conditions has many potential applications. These tasks are worth our expectations, but they are very challenging.

\section*{Acknowledgment}

This research was supported in part by the National Natural Science Foundation of China (Nos. 62002136 and 62272196), Natural Science Foundation of Guangdong Province (No. 2022A1515011861), Fundamental Research Funds for the Central Universities of Jinan University (No. 21622416), the Young Scholar Program of Pazhou Lab (No. PZL2021KF0023), Engineering Research Center of Trustworthy AI, Ministry of Education (Jinan University), and the Guangdong Key Laboratory of Data Security and Privacy Preserving.

\bibliographystyle{ACM-Reference-Format}
\bibliography{TALENT.bib}
\end{document}